\documentclass[aps,pre,twocolumn,showpacs,preprintnumbers,amsmath,amssymb,superscriptaddress]{revtex4-2}

\usepackage{graphicx}
\usepackage{dcolumn}
\usepackage{bm}
\usepackage{color}
\usepackage{hyperref}
\usepackage{amsmath}
\usepackage[capitalise]{cleveref}
\usepackage{amsfonts}
\usepackage{amsthm}
\usepackage{cases}
\usepackage{amssymb}
\usepackage{latexsym}
\usepackage{multirow}
\usepackage{xcolor}
\usepackage{mathrsfs}
\usepackage{textcomp}

\newcommand{\dd}{\mbox{d}}

\begin{document}
	\title{Bridging Wright-Fisher and Moran models}
	\date{\today}
	\author{Arthur Alexandre}
	\affiliation{Institute of Bioengineering, School of Life Sciences, \'Ecole Polytechnique F\'ed\'erale de Lausanne (EPFL), CH-1015 Lausanne, Switzerland}
	\affiliation{SIB Swiss Institute of Bioinformatics, CH-1015 Lausanne, Switzerland}
	
	\author{Alia Abbara}
	\affiliation{Institute of Bioengineering, School of Life Sciences, \'Ecole Polytechnique F\'ed\'erale de Lausanne (EPFL), CH-1015 Lausanne, Switzerland}
	\affiliation{SIB Swiss Institute of Bioinformatics, CH-1015 Lausanne, Switzerland}
	
	\author{Cecilia Fruet}
	\affiliation{Institute of Bioengineering, School of Life Sciences, \'Ecole Polytechnique F\'ed\'erale de Lausanne (EPFL), CH-1015 Lausanne, Switzerland}
	\affiliation{SIB Swiss Institute of Bioinformatics, CH-1015 Lausanne, Switzerland}
	
	\author{Claude Loverdo}
	\affiliation{Sorbonne Université, CNRS, Institut de
		Biologie Paris-Seine (IBPS), Laboratoire Jean Perrin (LJP), Paris, France}
	
	\author{Anne-Florence Bitbol}
	\affiliation{Institute of Bioengineering, School of Life Sciences, \'Ecole Polytechnique F\'ed\'erale de Lausanne (EPFL), CH-1015 Lausanne, Switzerland}
	\affiliation{SIB Swiss Institute of Bioinformatics, CH-1015 Lausanne, Switzerland}

	\date{\today}

	\begin{abstract}
The Wright-Fisher model and the Moran model are both widely used in population genetics. They describe the time evolution of the frequency of an allele in a well-mixed population with fixed size. We propose a simple and tractable model which bridges the Wright-Fisher and the Moran descriptions. We assume that a fixed fraction of the population is updated at each discrete time step. In this model, we determine the fixation probability of a mutant and its average fixation and extinction times, under the diffusion approximation. We further study the associated coalescent process, which converges to Kingman's coalescent, and we calculate effective population sizes. We generalize our model, first by taking into account fluctuating updated fractions or individual lifetimes, and then by incorporating selection on the lifetime as well as on the reproductive fitness.
	\end{abstract}
	
	\maketitle

	\section{Introduction}
	
	A major goal of population genetics is to describe how the frequency of an allele in a population changes over time. Different evolutionary forces shape genetic diversity. In the simplest models of well-mixed haploid populations with fixed size, the fate of a mutant is determined by the interplay of natural selection and genetic drift (in the absence of any additional mutation). The first model incorporating these ingredients dates back to the 1930s and is known as the Wright-Fisher model \cite{fisher1930genetical, wright1931evolution}. This widely used model assumes discrete and non-overlapping generations. Each generation is formed from the previous one through binomial sampling. This model 
	can be extended to include for instance multi-allele selection, additional mutations \cite{ewens1982concept} and is a suitable framework for the
	inference of selection from genetic data \cite{Tataru16, tataru2015inference, paris2019inference}. Another traditional model including natural selection and genetic drift is the Moran process. It considers overlapping generations with one individual birth and one individual death occurring at each discrete time step \cite{moran1958random}.
	Extensions of this model include other birth-death processes, e.g.\ in structured populations on graphs \cite{lieberman2005evolutionary,hindersin2015most, kaveh2015duality, marrec2021toward}. This framework is widely applied beyond population genetics, including evolutionary game theory \cite{taylor2004evolutionary, nowak2004emergence, ohtsuki2006replicator}. While the Wright-Fisher model and the Moran model share the same key ingredients and behave similarly, their different detailed dynamics lead to slightly different diffusion equations and mutant fixation probabilities. Specifically, in one generation, genetic drift is twice as large in the Moran model as in the Wright-Fisher model, while natural selection has the same effect. Hence, selection is more efficient in the Wright-Fisher model than in the Moran one, for instance in the fixation of beneficial mutations.
	
	Several population genetics models that are more general than the Wright-Fisher and Moran model have been developed. They include the Karlin and McGregor model \cite{karlin1964direct, karlin1965direct}, the Chia and Watterson model \cite{chia1969demographic}, and in the neutral case, the Cannings model~\cite{cannings1974latent}. The Chia and Watterson model considers a haploid population of fixed size $N$, where each individual reproduces independently, according to a specific distribution of offspring number. Then $r$ parents and $N-r$ offspring constitute the next generation, where $r$ is a random variable conditioned by the offspring population size. This encompasses Wright-Fisher and Moran models as particular cases. However, as pointed out by Cannings \cite{cannings1973equivalence}, the Chia and Watterson model is rather complex since it involves the offspring distribution, the fraction of surviving offsprings, and the sampling of the new generation from the parent and the offspring populations. It makes it hardly tractable, and to our knowledge, the mutant fixation probability has not been derived under this model. 
	
	Here, we consider a simple and tractable case of the Chia and Watterson model,  which bridges Wright-Fisher and Moran models. Specifically, we assume that a fraction of the population is updated at each discrete time step. 
	
	Such intermediate updated fractions are relevant in many biological situations. For instance, bacterial populations can comprise a fraction of slow-growing persister cells \cite{balaban_bacterial_2004,levy_bet_2012,fisher_peristent_2017,levin-reisman_antibiotic_2017}. Under stress, some bacteria can also undergo sporulation \cite{veening_bet-hedging_2008}. Such bet-hedging strategies, where a fraction of the population is dormant, facilitate population survival to harsh environmental conditions, and exist beyond bacteria. For instance, in annual plants, only a fraction of seeds may germinate each year, while others stay dormant \cite{sarukhan_studies_1977,ellner_ESS_1985,hyatt_isdecreased_1998,evans_germ_2005, venable_bet_2007, gremer_bet_2014}. 
	Besides, in a population where adults live for several years, with a reproduction season each year, and where mortality mainly arises from external factors such as infectious diseases or predation, the high mortality among young individuals results in an effective renewal of a fraction of the population during each reproductive season~\cite{sibly1997mortality,
		pike2008estimating}. 
	Seed banks, where seeds are preserved and can yield offspring many generations later, can also give rise to complex partial generation overlaps~\cite{epling1960breeding, kaj2001coalescent, shoemaker2018evolution}.

	To study this model, we focus on genetic drift and natural selection in a haploid population. We first assume that a fixed fraction of the population is updated at each discrete time step. Under Kimura's diffusion approximation, we obtain a simple expression of the fixation probability of a mutant, and determine the average conditional and unconditional absorption times. In the neutral case, we further study the rate of approach to the steady state, and the coalescent process associated to our model. This allows us to determine the variance effective population size, the eigenvalue effective population size, and the coalescent effective population size, which coincide. Besides, the ancestral process converges to Kingman's coalescent. Next, we generalize our model, first by taking into account fluctuations of the fraction that is updated at each discrete time step, and then by considering individuals with fluctuating lifetimes. This allows us to express the fixation probability as a function of the mean life time of individuals. Finally, we further generalize our model by incorporating selection on the lifetime as well as on the reproductive fitness, i.e.\ both on birth and death processes, assuming that lifetimes are geometrically distributed.

	\section{Model with fixed updated fraction}
	
	\subsection{Description of the model}
	
	Let us consider a population comprising a fixed number $N$ of haploid individuals. Let us assume that there are two types of individuals, corresponding to wild-type and mutant individuals. Let us denote by $1+s$ the relative fitness of the mutant compared to the wild-type. Let us further denote by $i$ the number of mutant individuals, which fully describes the state of the population, and let us introduce
	\begin{equation}
		x=\frac{i}{N}\,\,\,\,\textrm{and}\,\,\,\,p=\frac{i (1+s)}{i (1+s) + N - i}=\frac{x (1+s)}{1+x s}\,.\label{eq:xp}
	\end{equation}
	
\begin{figure}[ht!]
		\centering
		\includegraphics[width = \columnwidth]{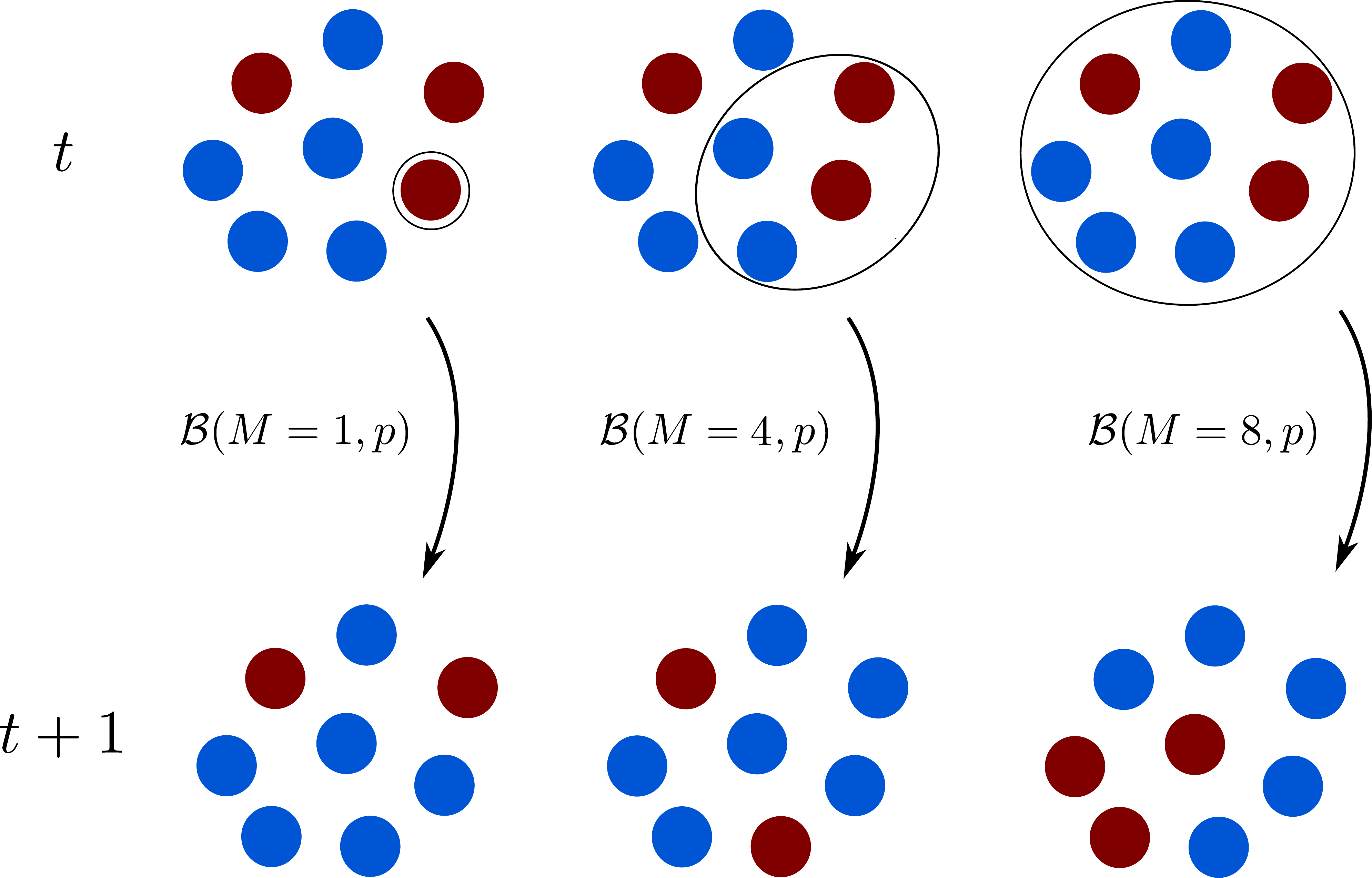}
		\caption{Schematic of three different updates in a population starting from $5$ wild-type individuals (blue) and $3$ mutants (red). Updates are performed by replacing $M$ individuals through binomial sampling with probability $p$ given in \cref{eq:xp}. Left: $M =1$ (which corresponds to the Moran model, see \cref{eq:Moran}), middle: $M=4$, right: $M=N=8$ (Wright-Fisher model). \label{fig:schematic}}
	\end{figure}

	At each discrete time step, we update a fraction of the  population as follows (see \cref{fig:schematic}):
	\begin{itemize}
		\item Choose $M$ individuals uniformly at random in the population, without replacement. $M$ satisfies $1\leq M\leq N$ (in this section, we assume that $M$ is fixed throughout the evolution of the population, and in the next sections, we will present generalizations).
		\item Replace the $M$ chosen individuals with $M$ new ones, sampled from a binomial distribution with parameters $M$ and $p$.
	\end{itemize}
	
	\paragraph*{Link to the Wright-Fisher model} For $M=N$, we recover the Wright-Fisher model. As a reminder, in the Wright-Fisher model, generations do not overlap. At each discrete step a new generation is sampled from the binomial law with parameters $N$ and $p$, yielding the following transition probabilities for the number of mutants:
	\begin{equation}
		P_{i \rightarrow j }^{\text{WF}(N)} = \binom{N}{j} {p}^j (1-p)^{N-j}\,,\label{eq:binomial_WF}
	\end{equation}
	for all $j$ between 0 and $N$. 
	
	\paragraph*{Formalization of the model} The number $k$ of mutants among the $M$ individuals sampled from the population follows a hypergeometric distribution with group sizes $i$ and $N-i$ and sample size $M$. Denoting by $q_{N,M,i}(k)$ the probability to draw $k$ mutants according to this distribution, we have:
	\begin{equation}
		q_{N,M,i}(k)  = \frac{\binom{i}{k} \binom{N-i}{M-k}}{\binom{N}{M}}\,.
	\end{equation}
	The transition probabilities for the number of mutants upon each discrete step then read:
	\begin{equation}
		P_{i \rightarrow j}^{(N,M)} = \sum_{k=0}^i q_{N, M, i}(k) P_{i\rightarrow k+j-i }^{\text{WF}(M)}\,, \label{eq:p_ij_eq}
	\end{equation}
	where $ P_{i \rightarrow j }^{\text{WF}(N)}$ is defined in \cref{eq:binomial_WF}.
	Note that for $M =N$, we have
	\begin{equation}
		q_{N,N,i}(k)  = \frac{\binom{i}{k} \binom{N-i}{N-k}}{\binom{N}{N}} = \delta_{k,i}\,,
	\end{equation}
	where $\delta_{i,j}$ is 1 if $i=j$ and 0 otherwise, and \cref{eq:p_ij_eq} reduces to \cref{eq:binomial_WF} as expected. Note also that the transition probabilities given in \cref{eq:p_ij_eq} are duly normalized. Indeed, introducing $m=k+j-i$, we have
	\begin{equation}
		\sum_{k=0}^i q_{N, M, i}(k) \sum_{m=0}^{M}P_{i\rightarrow m }^{\text{WF}(M)}=1\,.
	\end{equation}
	
	\paragraph*{Link to the Moran model} For $M = 1$, \cref{eq:p_ij_eq} can be simplified into 
	\begin{equation}
		P_{i \rightarrow j}^{(N,1)}= \left\{ \begin{array}{ll}
			x(1-p) & \text{if } j =i-1, \\
			x p + (1-x)(1-p) & \text{if }j =i, \\
			p(1-x) & \text{if }j =i +1, \\
			0 & \text{else,}
		\end{array} \right. \label{eq:Moran}
	\end{equation}
	which exactly coincides with the Moran model. 
	Therefore, our model generalizes over both the Wright-Fisher model and the Moran model. 
	
	\subsection{Mutant fixation probability}

Let us consider a large-sized population $N\gg 1$ and assume that $s$ is of order $1/N$. These conditions allow us to employ Kimura's diffusion approximation \cite{kimura1962probability,crow_introduction_1970,otto_biologists_2007} (see Supplementary Material for mathematical details). The fixation probability $\phi(x)$ of the mutant type, starting from a fraction $x$ of mutants in the population, then satisfies the following Kolmogorov backward equation \cite{kimura1962probability}:
\begin{equation}
	\frac{\mathbb{V}(\Delta x)}{2} \frac{\dd^2 \phi}{\dd x^2} +\mathbb{E}(\Delta x) \frac{\dd \phi}{\dd x} = 0\,, \label{eq:backward_FP}
\end{equation}
where $\Delta x$ is the variation of the mutant fraction occurring in the first discrete time step starting from the initial fraction $x$, and $\mathbb{E}(\Delta x)$ and $\mathbb{V}(\Delta x)$ denote respectively its expected value and variance. 
To leading order in $1/N$, they read (see Supplementary Material)
\begin{align}
	\mathbb{E}(\Delta x) & = \rho s x (1- x ) \,, \\
	\mathbb{V}(\Delta x) & =  \frac{\rho(2 - \rho)  }{N}  x (1- x ) \label{eq:varDeltax}\,,
\end{align}
where $\rho = M/N$ is the fraction of the population that is updated at each discrete time step. 

\begin{figure}[!ht]
		\centering
		\includegraphics[width = \columnwidth]{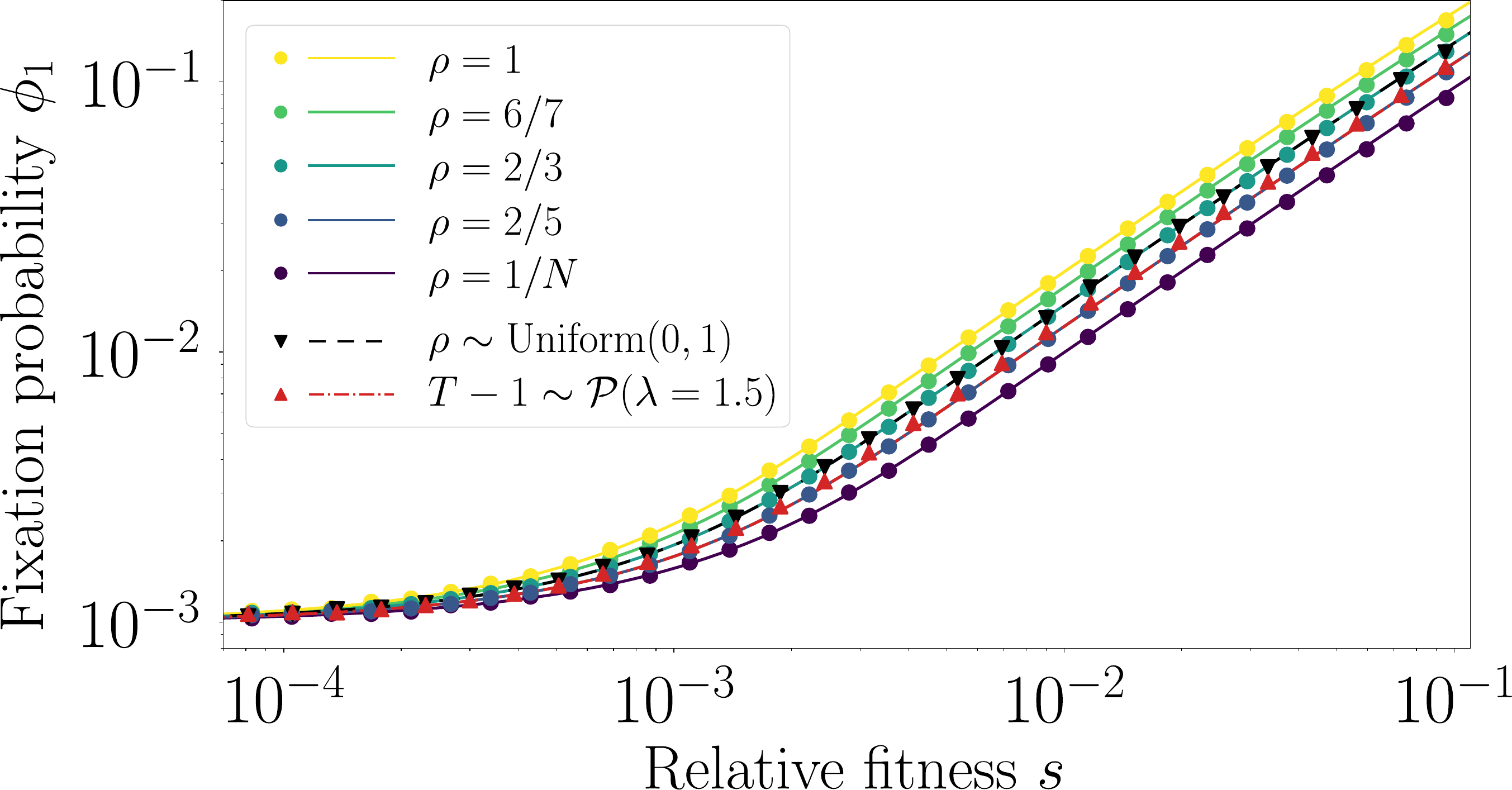}
		\caption{Fixation probability $\phi_1$ as a function of the mutant relative fitness advantage $s$ for $N=1000$. The dots correspond to numerical simulation results obtained from $10^7$ runs, and the lines to our theoretical predictions. Solid lines: constant sampling fraction $\rho$. Dashed black line: fluctuating sampling fraction drawn uniformly between $0$ and $1$ at each time step. Red dash-dotted line: fluctuating individual life time drawn from the law $T =  1+ X$ where $X$ follows a Poisson law with parameter $\lambda = 1.5$. }
		\label{fig:phi_1_plot}
	\end{figure}

Solving \cref{eq:backward_FP}, and using the boundary conditions  $ \phi(0) =0 , \ \phi(1) =1  $, 
we find that the fixation probability is 
\begin{equation}
	\phi(x) = \frac{1- \exp\left(-\frac{2 Nsx}{2 - \rho} \right)}{1- \exp\left(-\frac{2 Ns}{2 - \rho}\right)}\,.\label{eq:fixproba}
\end{equation}
This result generalizes over the fixation probability of both the Moran and the Wright-Fisher models in the diffusion regime. We recover them by taking $\rho=1/N$ (which is then neglected since we restrict to leading order in $1/N$), and $\rho = 1$, respectively. In \cref{fig:phi_1_plot}, we show the probability $\phi_1=\phi(1/N)$ of fixation in \cref{eq:fixproba} for different values of the updated fraction $\rho$. We observe that intermediate values of $\rho$ lead to probabilities of fixation that  are between those given by the Moran model and the Wright-Fisher model. The nonlinear dependence on $\rho$ of the fixation probability is illustrated in Figs.~S1 and S2 of the Supplementary Material.
	
Starting from one mutant, and assuming $s\ll 1$ while $Ns\gg 1$ and $N\gg 1$, \cref{eq:fixproba} yields to leading order in $1/N$
\begin{equation}
	\phi_1=\frac{2 s}{2 - \rho}\,.
\end{equation}
In this regime, the mutant probability of fixation increases with the updated fraction $\rho$ at fixed $s$, forming a continuum between the Moran result ($\phi_1=s$) and the Wright-Fisher one ($\phi_1=2s$). 
We note that this is analogous to Haldane's formula \cite{boenkost2021haldane_strong}
\begin{equation}
	\phi_1=\frac{2 s}{\sigma^2}\,,
\end{equation}
where $\sigma^2 = 2 - \rho$ is the variance of the number of offspring drawn from a Poisson distribution (see Supplementary Material).

	\begin{figure*}[t]
	\centering
	\includegraphics[width = 0.8\linewidth]{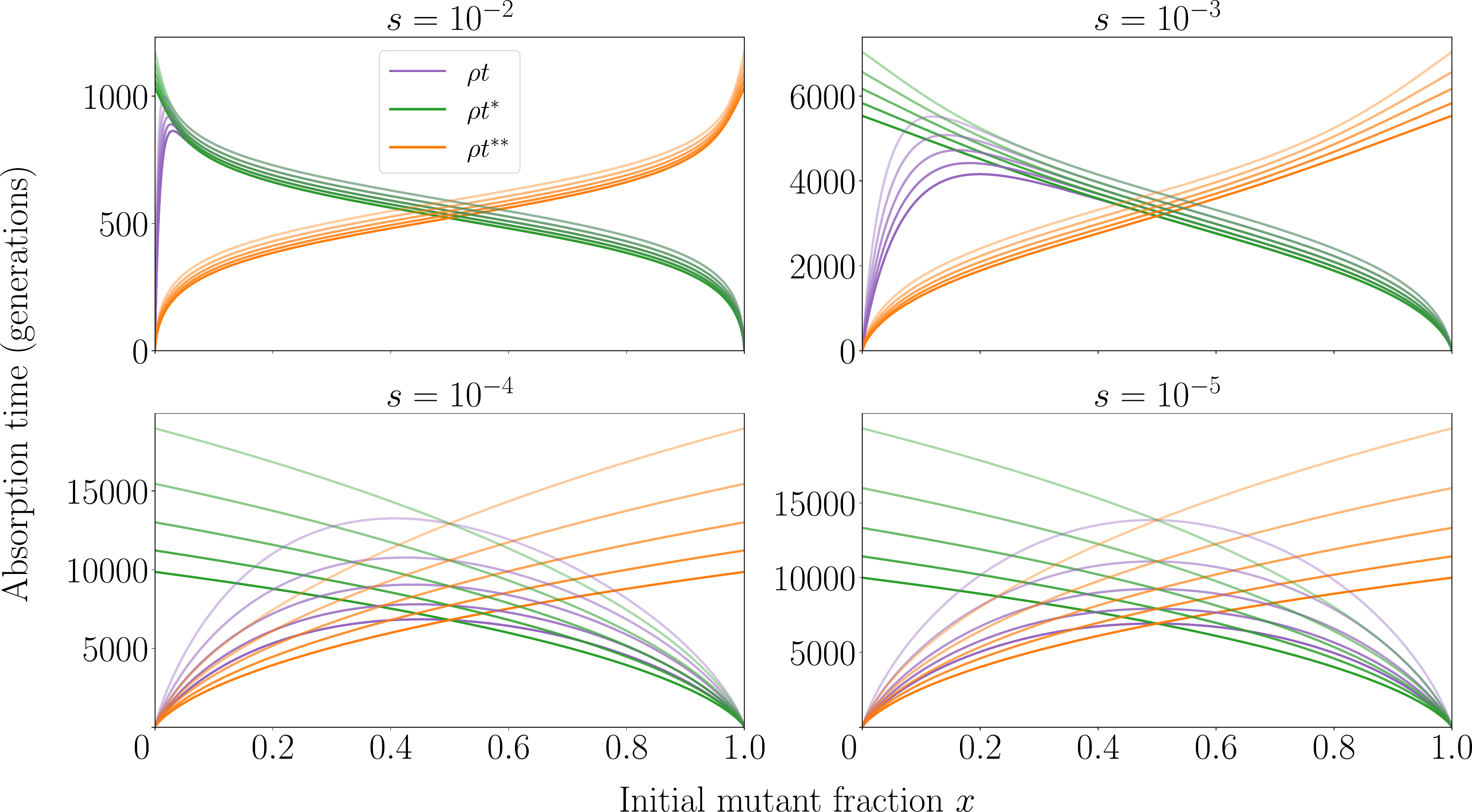}
	\caption{The mean unconditional absorption time $t$, as well as the mean mutant fixation time $t^{*}$ and extinction time $t^{**}$ are shown in generations (i.e.\ rescaled with the generation time $1/\rho$) versus the initial mutant fraction $x$, for $N=10^5$. Each of these times is plotted for $\rho = 1/N, 0.25, 0.5, 0.75, 1$, where increasing values of $\rho$ correspond to lighter colors. Different panels show different relative mutant fitness advantages $s$. }
	\label{fig:t_fixation_plot}
\end{figure*}

	\subsection{Mean absorption times}
	Kimura's diffusion approximation also allows determining the mean absorption times, i.e.\ the time it takes for the mutant type to either fix in the population or to get extinct. Note that in the long term, fixation or extinction are the only possible outcomes: they are the two absorbing states of the system. 
	
	The mean unconditional absorption time $t(x)$ to reach either fixation or extinction, starting from an initial fraction $x$ of mutants with relative fitness $1+s$, satisfies~\cite{ewens2004mathematical}
	\begin{equation}
		\frac{\mathbb{V}(\Delta x)}{2} \frac{\dd^2 t}{\dd x^2} +\mathbb{E}(\Delta x) \frac{\dd t}{\dd x} = -1\,, \label{eq:backward_time}
	\end{equation}
	and $t(0) = t(1) =0$. In the Supplementary Material, we provide integral solutions to this equation.
	
	The mean absorption time $t^{*}$ conditioned on mutant fixation, and $t^{**}$ conditioned on mutant extinction, can be obtained in a similar way~\cite{ewens2004mathematical} (see Supplementary Material).
	
	These three absorption times, rescaled to be expressed in generations, are shown in \cref{fig:t_fixation_plot} versus the initial mutant fraction $x$ for different values of $\rho$ (different shades) and $s$ (different panels). We observe that increasing the updated fraction $\rho$ gradually increases all three absorption times, once they are expressed in generations. Besides, when the relative fitness advantage $s $ of the mutant decreases and goes toward effective neutrality, absorption times increase, and the unconditional one, $t$, becomes more symmetric around $x=1/2$, as for Moran and Wright-Fisher models. In addition, for $s>1/N$, the unconditional absorption time $t$ is close to the fixation time $t^{*}$ for sufficiently large $x$, as fixation then dominates. We further show in the Supplementary Material the symmetry property $t^{*}(1-x) = t^{**}(x)$.
	
	\subsection{Rate of approach to steady state in the neutral case}
	
	In this section, we consider neutral mutants, i.e.\ $s = 0$. The Markov chain representing the time evolution of the number of mutants in the population possesses two absorbing states, namely fixation and extinction of the mutant type. Therefore, the transition matrix in our model, whose elements are given by \cref{eq:p_ij_eq}, has two unit eigenvalues \cite{ewens2004mathematical}. The next largest non-unit eigenvalue provides information on the rate of decrease of genetic diversity in the population. It is thus of interest to calculate it. Denoting by $X_t$ be the number of mutant at time $t$, we can show that (see Supplementary Material)
	\begin{align}
		\mathbb{E}\left[X_{t+1}(N-X_{t+1}) | X_t\right] & = X_{t}(N-X_{t}) \lambda_*\,, \label{eq:rec_lambda_star}
	\end{align}
	where, to leading order in $1/N$, we have:
	\begin{align}
		\lambda_* = 1- \frac{\rho (2- \rho)}{N}\,.
		\label{eq:lambda_star}
	\end{align}
	A theorem detailed in Appendix A of Ref.~\cite{ewens2004mathematical} ensures that $ \lambda_*$ is the highest non-unit eigenvalue of the transition matrix of the model.

	The rate of approach to steady state is $\lambda_*$ for one event, i.e.\ for one discrete time step of the model \cite{moran1958random}. Thus, for one generation, which corresponds to $1/\rho$ discrete time steps, it is, still to leading order in $1/N$,
	\begin{equation}
		r = \lambda_*^{1/\rho} = 1- \frac{2-\rho }{N} \,.\label{eq:rate}
	\end{equation}
	Again, we recover both Moran and Wright-Fisher limits by taking $\rho=1/N$ and $\rho=1$, respectively \cite{moran1958random, cannings1974latent}. 
	
	The convergence to steady state can be quantified by considering the vector $\boldsymbol{\pi}_{t} = (p_0(t),p_1(t),p_2(t),..., p_N(t))$ where $p_k(t)$ denotes the probability of having $k$ mutants at time $t$ starting from $1$ mutant at $t =0$. For $t\rightarrow \infty$, we have (see Supplementary Material)
	\begin{equation}
		||\boldsymbol{\pi}_t - \boldsymbol{\pi}_{\infty} ||_{2}\propto \lambda_{*}^t = {r}^{\rho t} \,,
		\label{eq:convergence_markov_main}
	\end{equation}
	where $||.||_{2}$ refers to the Euclidean norm and 
	\begin{equation}
		\boldsymbol{\pi}_{\infty} = \lim\limits_{t\rightarrow\infty}\boldsymbol{\pi}_{t} = (1-1/N,0, 0, ...,0, 1/N).
	\end{equation}
	
\begin{figure}[t]
		\centering
		\includegraphics[width = \columnwidth]{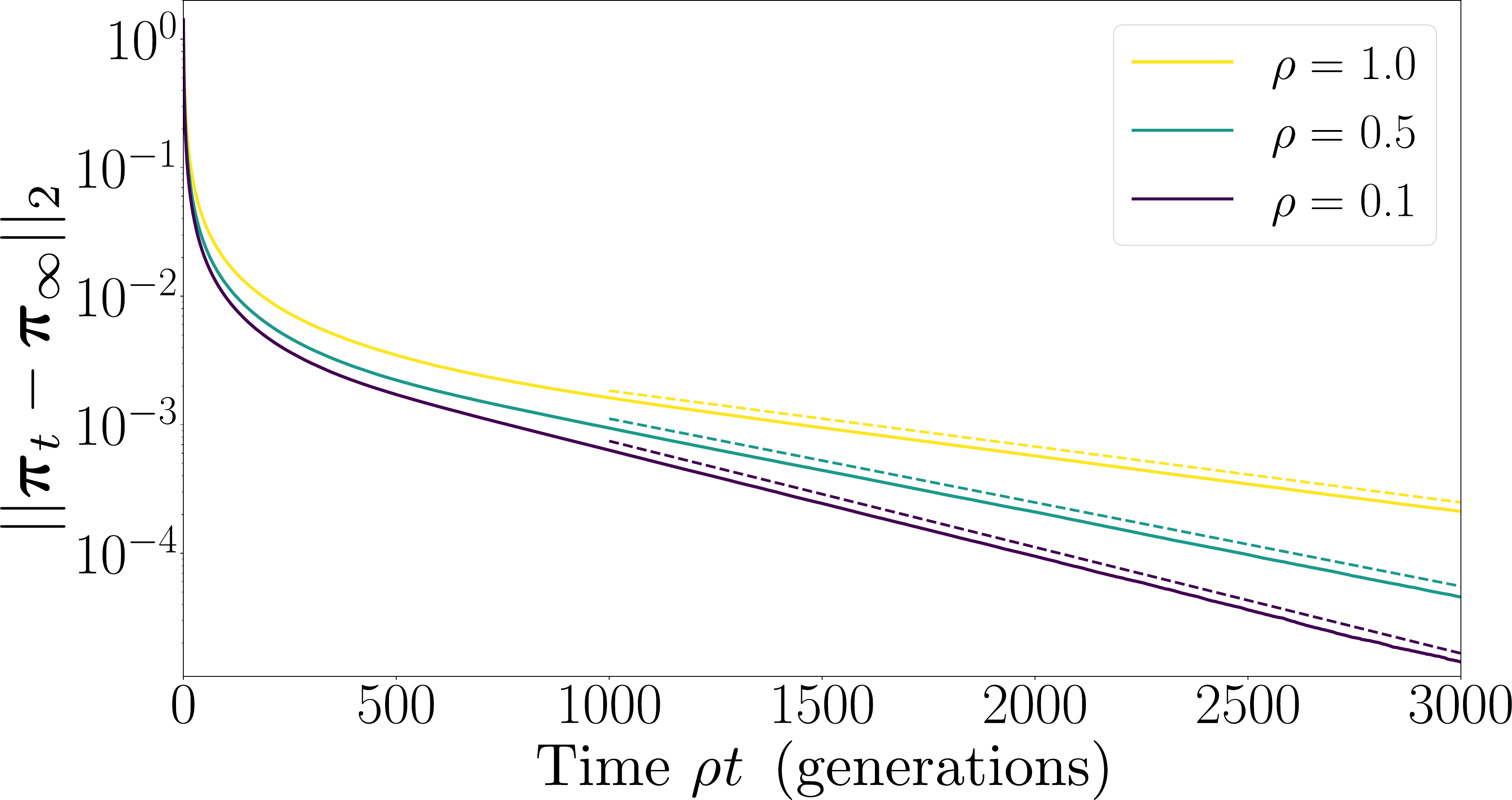}
		\caption{Approach to steady state in the neutral case. We show the Euclidean norm of $\boldsymbol{\pi}_t-\boldsymbol{\pi}_{\infty} $ (see main text) versus time in generations (i.e.\ rescaled with the generation time $1/\rho$), for $N = 1000$ and with different updated fractions $\rho$. Full lines: simulation results obtained from $5.10^{7}$ runs. Dashed lines: asymptotic behaviour given in \cref{eq:convergence_markov_main}, with a multiplicative constant chosen for legibility. }
		\label{fig:lambda_star_plot}
	\end{figure}

	In \cref{fig:lambda_star_plot}, this prediction is tested numerically for different updated fractions $\rho$ of the population. We find a good agreement between our simulation results and the long-time asymptotic behaviour given in \cref{eq:convergence_markov_main}. Increasing the updated fraction $\rho$ increases the time it takes to approach steady state, measured in generations. Again, our model leads to a continuum between the Moran model and the Wright-Fisher model.
	
The eigenvalue in \cref{eq:lambda_star} is related to the expected diversity $H(t)$ of the population (also called expected heterozygosity, especially in diploid populations), defined as the probability of drawing two individuals of different types (here, wild-type and mutant). Indeed, it reads:
\begin{equation}
	H(t) = \frac{\mathbb{E}\left[ X_t(N-X_t)\right]}{\binom{N}{2}} = \lambda_*^t H(0) ,
\end{equation}
where we have used the derivation of \cref{eq:rec_lambda_star} (see Supplementary Material). Thus, $\lambda_*$ also characterizes the rate of loss of diversity in the population.

\subsection{Coalescent}
The coalescent traces the genealogy of a given sample of individuals backward in time to their most recent common ancestor. A large class of models converge to Kingman's coalescent \cite{kingman1982a, kingman1982b, hudson1983, tajima1983} for neutral mutants in the limit of large population size. This includes Moran and Wright-Fisher models, but also the broad class of Cannings models~\cite{mohle1998,mohle2001classification}. In the specific case of neutral mutants, our partial update model falls in the latter class. Therefore, under the proper time scaling, the coalescent process associated with our model in the neutral case also converges to Kingman's coalescent in the limit of large populations~\cite{kingman1982a, mohle2001classification, SARGSYAN2008104}, see Supplementary Material for details. 

For completeness, let us also explicitly consider our model for finite population size $N$, still in the neutral case. A central quantity in coalescent processes is the probability $G_{i,j}$ that $i$ sampled individuals at discrete time step $t$ have exactly $j$ parents at $t-1$. In our model, we show in the Supplementary Material that this probability reads
\begin{align}
	G_{i,j} = \sum_{k=i-j}^{\min(i, M)} \frac{\binom{N-M}{i-k}\binom{M}{k}}{\binom{N}{i}} \frac{(N-i+k)!}{(N-j)!N^k} {\left\{\genfrac{}{}{0pt}{}{i}{j}  \right\}}_{i-k},
\end{align}
where ${\left\{\genfrac{}{}{0pt}{}{i}{j}  \right\}}_{r}$ represents the $r$-Stirling number of the second kind~\cite{broder1984r}. In the large-$N$ limit, and taking $i \ll N$, this yields  
(see Supplementary Material)
\begin{align}
	G_{i,i} &=1- \binom{i}{2} \frac{\rho(2-\rho)}{N} + o\left(\frac{\rho}{N} \right), \label{G1}\\
	G_{i,i-1} &= \binom{i}{2} \frac{\rho(2-\rho)}{N} + o\left(\frac{\rho}{N} \right). \label{G2}
\end{align}
Here too, taking $\rho =1/N$ and $\rho =1$ yields respectively the Moran model result and the Wright-Fisher model result~\cite{gladstien1978}. 
These expressions also illustrate the link with Kingman's coalescent (see Supplementary Material).

Furthermore, \cref{G1,G2} can be used to express the mean coalescence time, i.e.\ 
the mean time $T_{\text{MRCA}}$ to the most recent common ancestor of the population, for finite $N$ (see Supplementary Material). 
In the large-$N$ limit, the mean coalescence time reads~\cite{ewens2004mathematical, tavare2004part} 
\begin{equation}
	T_{\text{MRCA}} = \frac{2(N-1)}{\rho (2-\rho)}.
	\label{eq:tmrcalargeN}
\end{equation}
Note that this time is expressed in number of discrete time steps -- thus, $\rho T_{\text{MRCA}}$ represents the coalescence time expressed in generations. The mean coalescence time $\rho T_{\text{MRCA}}$ is shown versus population size $N$ for various values of $\rho$ in Fig.~S3 of the Supplementary Material, and the large-$N$ expression in \cref{eq:tmrcalargeN} is observed to be a good approximation even for moderate $N$.

\subsection{Effective population sizes}
\label{sec:effective_pop}
The effective population size is the size a regular Wright-Fisher population should have to match a given property of the population considered. Several definitions can be found in the literature, each one matching a different property of the Wright-Fisher population, see~\cite{ewens2004mathematical, orive1993effective} for a review.  Here, we determine several of these effective population sizes for a population of size $N$ within our partial update model. 

Let us first consider the variance effective population size $N_{e,v}$. We can determine it in our model starting from the definition in Refs.~\cite{ewens1982concept,ewens2004mathematical}:
\begin{equation}
	N_{e,v} \equiv \frac{ x(1-x)}{\mathbb{V}_g(\Delta x) } = \frac{ x(1-x)}{\mathbb{V}(\Delta x) / \rho}= \frac{N}{2-\rho}.
\end{equation}
Here, we denoted by $\mathbb{V}_g$ the variance accumulated over one generation~\cite{ewens2004mathematical}. Because the average generation time is $1/\rho$ in our model, $\mathbb{V}_g(\Delta x)$ is $1/\rho$ times the variance in \cref{eq:varDeltax}, which allowed us to conclude. 

Next, let us consider the eigenvalue effective population size $N_{e,\lambda}$. \cref{eq:rate} allows us to express it as \cite{ewens1982concept}
\begin{equation}
	N_{e,\lambda} = \frac{1}{1-r} = \frac{N}{2-\rho}\,.
\end{equation}

Finally, since our model converges to Kingman's coalescent for large $N$ in the neutral case with a proper time rescaling, we can define a coalescent effective size $N_{e,c}$~\cite{nordborg2002separation, sjodin2005meaning, wakeley2009extensions}. Note that this definition of the effective population size is sometimes considered more general than others, as a large variety of models converge to Kingman's coalescent~\cite{kingman1982a, mohle2001classification}. Besides, the coalescent essentially contains all of the information that can be found in sampled genetic data~\cite{nordborg2002separation, sjodin2005meaning, wakeley2009extensions}. We adopt the most general definition given in Ref.~\cite{wakeley2009extensions}, which depends on the properties of the ancestral process, as well as on the time scaling (see Supplementary Material). The resulting effective coalescent size is
\begin{equation}
	N_{e,c}= \dfrac{N}{2 - \rho}.
\end{equation}

The three definitions of effective population size coincide for our model, giving a unique expression $N_e$. 
Fig.~S4 in the Supplementary Material illustrates how the effective population size $N_e$ increases non-linearly with the updated fraction $\rho$, from the Moran result, $N_e=N/2$, to the Wright-Fisher one, $N_e=N$. Populations with partially-overlapping generations have intermediate effective population sizes. 
Furthermore, the effective size $N_e$ is a convex function of the updated fraction $\rho$ (as its derivative with respect to $\rho$ grows with $\rho$). Thus, most intermediate values of $\rho$ yield effective population sizes closer to the Moran result than to the Wright-Fisher one. 
	
	\section{Generalizations of the model}
	
	\subsection{Fluctuating updated fraction }
	Let us extend our model where a fraction of the population is updated at each discrete time step to the case where the number $M$ of individuals selected to be updated can fluctuate. Let us assume that there is a probability $\pi^{(N)}(M)$ to have each value of $M$ between 0 and $N$. In a discrete time step of this new model, the transition probability is
	\begin{equation}
		P_{i \rightarrow j}^{\left(N,\pi^{(N)}\right)} = \sum_{M=0}^N \pi^{(N)}(M) P_{i \rightarrow j}^{(N,M)}\,, 
	\end{equation}
	where $P_{i \rightarrow j}^{(N,M)}$ is given by \cref{eq:p_ij_eq}. To obtain the mutant fixation probability in the diffusion approximation within this new model, we express the first two moments of $\Delta x$. The calculation is similar to the one presented in the Supplementary Material for the model with fixed updated fraction, and yields to leading order in $1/N$
	\begin{align}
		\mathbb{E}(\Delta x) &= \left< \rho \right> s x(1- x )\,,\\
		\mathbb{V}(\Delta x) &=  \frac{ \left<\rho (2 - \rho)\right>}{N}   x(1 - x )\,,
	\end{align}
	where $\langle . \rangle$ denotes the average under the distribution $\pi^{(N)}$. Thus, the mutant fixation probability reads
	\begin{equation}
		\phi(x) = \frac{1- \exp\left(-\frac{2 Nsx}{2 - \langle\rho^2\rangle/\langle\rho\rangle} \right)}{1- \exp\left(-\frac{2 Ns}{2 - \langle\rho^2\rangle/\langle\rho\rangle}\right)}\,.\label{eq:fixproba2}
	\end{equation}
	Starting from one mutant, and assuming $s\ll 1$ while $Ns\gg 1$ and $N\gg 1$, \cref{eq:fixproba2} yields to leading order in $1/N$
	\begin{equation}
		\phi_1=\frac{2 s}{2 - \left< \rho^2\right>/\left< \rho \right>} \,.\label{eq:phi1rho}
	\end{equation}

	For instance, if $\pi^{(N)}$ is the uniform distribution between 0 and $N$, then we obtain $\phi_1= 3 s / 2$ in this regime, which is the same prefactor as the one found for a constant update fraction $\rho = 2/3$ (see \cref{fig:phi_1_plot}). Furthermore, we note that 
	\begin{equation}
		1\leq \frac{2 }{2 - \left< \rho^2\right>/\left< \rho \right>} \leq 2.
	\end{equation}
	The first inequality follows from $\langle \rho^2\rangle \geq 0$. Moreover, since $\forall\rho \in \left[0,1\right]$, $\rho^2\leq \rho$, we have $\langle \rho^2\rangle \leq \langle \rho\rangle$ for any $\pi^{(N)}$, yielding the second inequality. This result entails that the fixation probabilities in this model are comprised between those of the Moran and of the Wright-Fisher models.
	
	Let us determine how $\rho$ is related to the lifetime of individuals $T$, i.e.\ to the number of time steps between two successive updates involving one given individual. The probability that a specific individual is selected to be updated at a given discrete time step is given by $M/N$, and $M$ is distributed according to $\pi^{(N)}$. Then, the probability that $T$ is equal to $k$ time steps, where $k$ is a positive integer, can be written as the probability that the individual of interest is not chosen at any of the $k-1$ previous updates and then is chosen, i.e.\
	\begin{align}
		p(T=k) &= \left[\sum_{M=0}^N \pi^{(N)}(M) \frac{N-M}{N} \right]^{k-1} \sum_{M=0}^N \pi^{(N)}(M) \frac{M}{N} \nonumber\\
		&=\left[1-\langle\rho\rangle\right]^{k-1}\langle\rho\rangle\,.
	\end{align}
	Thus, for any distribution $\pi^{(N)}$ of the fraction of the population chosen to be updated, the lifetime $T$ of an individual follows a geometric law with mean 
	\begin{align}
		\langle T\rangle& = 1 / \langle \rho\rangle .\label{eq:mean_T}
	\end{align}
	
	\subsection{Fluctuating lifetimes}
	
	Let us now consider a different model, which is specified through the distribution $\pi_\ell$ of lifetimes of individuals. As before, the lifetime $T$ of an individual is intended as the time between two successive updates involving this individual. Time is still discrete, and thus $\pi_\ell(T=k)$ is defined over positive integers $k$. We assume that each of the $N$ individuals has the same distribution $\pi_\ell$ of lifetimes, and that they are all independent. In particular, mutants and wild-types have the same distribution $\pi_\ell$ of lifetimes. We further assume that there is no memory: a given lifetime is independent of previous ones, and all are identically distributed according to $\pi_\ell$. 
	To relate this model to the previous ones, we aim to determine the distribution of the updated fraction $\rho$ from the lifetime distribution. Let us focus on the probability $p_u(t)$ for an individual to be updated at time $t$, which satisfies the equation
	\begin{equation}
		p_u(t) = \sum_{k= 1}^{t}p_u(t- k) \pi_\ell(T = k) \, ,
	\end{equation}
	with the initial condition $ p_u(t = 0) = 1 $.
	Note that if $\pi_\ell$ is a geometric distribution, then its parameter is $1/\langle T\rangle$, and we obtain
	$ p_u(t) = 1/\left<T\right>$, which does not depend on $t$. This result is not always valid for other distributions. However, renewal process theory shows that it holds in the limit $t\rightarrow+\infty$ for many distributions~\cite{erdos1949property, feller1949fluctuation}:
	\begin{equation}
		\lim\limits_{t\rightarrow+\infty}p_u(t)=1/\langle T\rangle\,.
	\end{equation}
	In this long-time limit, the distribution $\pi^{(N)}$ of the number $M$ of updated individuals is a binomial distribution $\mathcal{B}(N, p =1/\langle T\rangle )$, which leads to
	\begin{equation}
		\langle \rho\rangle=1/\langle T\rangle \,,
	\end{equation}
	and we find to leading order in $1/N$
	\begin{equation}
		\langle \rho^2\rangle/\langle \rho\rangle=1/\langle T\rangle\,.
	\end{equation}
	\cref{eq:phi1rho} then becomes
	\begin{equation}
		\phi_1=\frac{2 s}{2 - 1/\left< T \right>} \,.\label{eq:phi1rhob}
	\end{equation}
	Again, this is consistent with Wright-Fisher and Moran models, indeed we have $\langle T\rangle = 1$ in the Wright-Fisher model and $\langle T\rangle= N$ in the Moran model.

	\subsection{Selection on the lifetime}
	
	Let us now consider the case where mutants and wild-types have a different lifetime distribution. For simplicity, we start by neglecting selection on division, meaning that the Wright-Fisher update will be performed without selection, using the fraction $x$ of mutants and not its rescaled version $p$ (see \cref{eq:xp}). Concretely, let us assume that the lifetime of wild-types (resp.\ mutants) follows a geometric distribution with parameter $p_w$ (resp.\ $p_m = (1-s) p_w$). Note that $s$ still denotes the selection parameter, but it represents selection on lifetime here.
	Within this model, the transition probability can be written as
	\begin{equation}\begin{split}
			&P_{i\rightarrow j}^{(N,p_m,p_w)} = \sum_{M=0}^N\sum_{k=0}^i \binom{i}{k} p_m^k(1-p_m)^{i-k}\\&  \quad \quad \quad \quad \times \binom{N-i}{M-k} p_w^{M-k}(1-p_w)^{N-i-M+k}  \\& \quad  \quad \quad \quad\times\binom{M}{k+ j-i}x^{k+j-i} (1-x)^{M-k-j+i}\,.
		\end{split}
		\label{eq:P_i_j_selection_life_time}
	\end{equation}
	Indeed, each possible number $M$ of individuals chosen to be updated can be obtained by choosing $k$ mutants among $i$ according to their lifetime distribution, which is binomial (see above), and choosing $M-k$ wild-types among $N-i$ in a similar way. Finally, the Wright-Fisher update is performed for these $M$ chosen individuals, without selection.
	
	Assuming as usual that $s$ is of order $1/N$, we obtain, to leading order in $1/N$ (see Supplementary Material):
	\begin{align}
		\mathbb{E}(\Delta x) & = p_w s x (1-x)   \,, \\ \mathbb{V}(\Delta x) & = \frac{p_w(2- p_w)}{N} x(1-x)   \,.
	\end{align}
	This allows us to express the fixation probability in the diffusion approximation as in other cases. In particular, assuming $s\ll 1$ while $Ns\gg 1$ and $N\gg 1$ it yields, to leading order in $1/N$, 
	\begin{equation}
		\phi_1= \frac{2 s}{2 - p_w}= \frac{2 s}{2 - 1/\langle T\rangle} \,,
	\end{equation}
	where we used the fact that $p_w=1/\langle T\rangle$, with $\langle T\rangle$ the mean lifetime of individuals.
	Hence, the result takes the same form for selection on lifetime (death) as for selection on division (birth).
	
	It is possible to generalize further by considering selection both on birth and death events. For this, we write $p_m = (1- s_d) p_w$ and we perform Wright-Fisher sampling with parameters $m$ and $p = x (1+ s_b)/(1+ x s_b)$. Both $s_d$ and $s_b$ are assumed to be of order $1/ N$. This case can be treated as above, yielding the fixation probability in the diffusion approximation. In particular, assuming $s_d\ll 1$ and $s_b\ll 1$ while $Ns_d\gg 1$, $Ns_b\gg 1$ and $N\gg 1$, it yields, to leading order in $1/N$,
	\begin{equation}
		\phi_1= \frac{2 (s_b + s_d)}{2 - p_w}\,.\label{eq:bothsel}
	\end{equation} 
	Note that for a Moran process with selection both on birth and death, the mutant fixation probability reads~\cite{kaveh2015duality}
	\begin{equation}
		\phi_1 = \dfrac{1- \dfrac{1 - s_d}{1+ s_b} }{1- \left(\dfrac{1 - s_d}{1+ s_b} \right)^N}\,.
	\end{equation}
	In particular, assuming $s_d\ll 1$ and $s_b\ll 1$ while $Ns_d\gg 1$, $Ns_b\gg 1$ and $N\gg 1$, it yields, to leading order in $1/N$,
	\begin{equation}
		\phi_1 = s_b + s_d \,,
	\end{equation}
	which is consistent with \cref{eq:bothsel} for $p_w=1/N$.
	
	\section{Discussion}
We proposed a simple and tractable model that bridges the Wright-Fisher model and the Moran model, and can be seen as a particular case of the Chia and Watterson model~\cite{chia1969demographic}. We do not focus on the offspring distribution, but on the fraction $\rho$ of the population is updated at each discrete time step, using binomial sampling. This leads to a more tractable model where the Wright-Fisher and the Moran model emerge as simple limiting cases, respectively for $\rho=1$ and $\rho=1/N$. In our model, one generation corresponds to $1/\rho$ discrete time steps. We  showed that the rescaled process, taking one unit of time as $N/\rho$ discrete steps, converges to a diffusion. Under the diffusion approximation, we obtained a simple expression of the mutant fixation probability, and expressed the conditional and unconditional mean absorption times. We also calculated the rate of approach to steady state. We further studied the genealogy of a population sample in the neutral case, which converges to Kingman's coalescent for large population sizes~\cite{kingman1982a,mohle1998}. 
Convergence to the coalescent is achieved when taking one unit of coalescent time as $N/[\rho(2-\rho)]$ discrete time steps. This time scaling differs by a multiplicative constant from the one commonly used to show convergence to a diffusion process. We determined the variance, eigenvalue and coalescent effective population sizes in our model, and showed that they coincide. Throughout, we wrote key quantities characterizing the model as simple functions of the updated fraction $\rho$, and discussed its impact on these quantities. 

We extended our model to the case where the updated fraction $\rho$ fluctuates. We showed that the mean individual lifetime is the inverse of the mean updated fraction. We considered the case where the lifetime fluctuates. Finally, we examined the situation where selection occurs both on birth and on death, assuming that the lifetime of individuals is geometrically distributed. In all these cases, we obtained simple expressions of the fixation probability in the diffusion approximation.

Here, we did not model mutations, as we primarily focused on fixation probabilities and absorption times. Including mutations explicitly in our model would be an interesting extension. It would allow to derive the stationary distribution of mutants and wild-types in the population, since mutant fixation and extinction would no longer be absorbing states~\cite{ewens2004mathematical}. Previous works have investigated the impact of mutations on the effective coalescent size~\cite{wakeley2009extensions}, as well as the diffusion approximation~\cite{tavare1984}.
In the case of neutral mutants, our partial update model falls in the class of Cannings models~\cite{cannings1973equivalence,cannings1974latent}, which yield convergence to Kingman's coalescent~\cite{kingman1982a, kingman1982b, mohle1998, mohle2003coalescent} in the large-$N$ limit, even when parent-independent mutations are included. Furthermore, the stationary distribution of allele frequencies across generations is then the same as in the Wright-Fisher model~\cite{gan2017dirichlet}.

Including fluctuations in the total population size in addition to fluctuations of the updated fraction would account for ``sweepstakes effect'' observed for marine organisms, where adults release a large number of gametes. While most gametes are lost to external perturbations, reproduction is a chance event that can produce a very large number of offspring~\cite{beckenbach1994mitochondrial, hedgecock1994does}.

Here, we assumed discrete time steps, as in most classic models of population genetics. This is relevant in many situations, e.g.\ annual reproduction. Moreover, when continuous time is more relevant, as in bacterial reproduction, mutant replications remain synchronized for some time, as they stem from the same ancestor \cite{jafarpour2023evolutionary}. Besides, while we focused on haploids, our model could also be extended to diploids, and various hypotheses on reproduction could be investigated. 

Finally, we focused on a well-mixed population with fixed size, and we considered two types of individuals. It would be interesting to extend our model to spatially structured populations \cite{lieberman2005evolutionary,Otwinowski14, marrec2021toward,Abbara23}, or to fluctuations of the population size \cite{engen2005effective}, and to consider multiple sites and alleles \cite{ewens1982concept}. For instance, our study of the coalescent and of the effective coalescent size could be used to determine the fixation index in structured populations~\cite{crow_introduction_1970}, and the allele frequency spectrum in cases with multiple sites and alleles~\cite{ewens2004mathematical}, for populations with partial updates.

	\section*{Acknowledgments}
	A.~A. thanks Félix Foutel-Rodier for useful suggestions. This research was partly funded by the Swiss National Science Foundation (SNSF) (grant No.~315230\_208196, to A.-F.~B.), by the Chan-Zuckerberg Initiative (CZI, to A.-F.~B.), and by the European Research Council (ERC) under the European Union’s Horizon 2020 research and innovation programme (grant agreement No.~851173, to A.-F.~B.). C.~L. acknowledges funding by the Agence Nationale de la Recherche (grants ANR-20-CE30-0001 and ANR-21-CE45-0015, to C.~L.).

\widetext	
\clearpage

\begin{center}
	\textbf{\large Supplementary Material} 
	\vspace{0.4cm}
	
	\text{Arthur Alexandre, Alia Abbara, Cecilia Fruet, Claude Loverdo, Anne-Florence Bitbol}
\end{center}

\setcounter{equation}{0}
\setcounter{section}{0}
\setcounter{figure}{0}
\setcounter{table}{0}
\makeatletter
\renewcommand{\theequation}{S\arabic{equation}}
\renewcommand{\thefigure}{S\arabic{figure}}

\section{Diffusion approximation}
\label{sec:diffusion_approx}
\subsection{Simplified theorem from Ref.~\cite{ethier1980diffusion}}

Justifying the diffusion approximation for population genetics models is a challenging topic in probability theory. The diffusion approximation was introduced in a seminal work by Kimura~\cite{kimura1962probability} and further formalized by Ethier and Nagylaki~\cite{ethier1980diffusion}. In Ref.~\cite{ethier1980diffusion}, they considered Markov chains with two time scales. Here, we are interested in haploid individuals in a well-mixed population. Therefore, only one relevant timescale is involved, and we only need a simplified version of the theorem from Ref.~\cite{ethier1980diffusion}, that we state below.

Let $ x_t$ ($t=0,1,\dots $) be a homogeneous Markov chain in a metric space $E_N$ (where $N$ is the population size). Now let $\epsilon_N$ be a \textit{time scaling} such that  $\lim\limits_{N \rightarrow +\infty} \epsilon_N = 0$. 
Let $x=x_{0}$ be the value of the initial state of the Markov chain. Assuming that the Markov chain has a transition function such that, $\forall x \in E_N$,
\begin{itemize}
	\item[1)] $\epsilon_N^{-1} \mathbb{E} \left(\Delta x\right)=b(x)+o(1)$,
	\item[2)] $\epsilon_N^{-1} \mathbb{E} \left(\Delta x^2\right)=a(x)+o(1)$,
	\item[3)] $\epsilon_N^{-1} \mathbb{E} \left(\Delta x^4\right) = o(1)$.
\end{itemize}
where $\Delta x = x_{1}-x_{0}$. Then, provided some regularity assumptions on functions $a$ and $b$, $x_{[t/\epsilon_N]}$ 
(i.e.\ $x_t$ taken every $1/\epsilon_N$ steps) converges weakly to the diffusion process $x(\cdot)$ with generator:
\begin{equation}
	\mathcal{L}=\dfrac{a(x)}{2}\dfrac{\partial^2}{\partial x^2} + b(x) \dfrac{\partial}{\partial x}.
\end{equation}

To apply this theorem to simple population genetics models, we need to write the mean and variance of the variation of the fraction of mutants, for one step of the Markov chain. Rescaling those functions by the appropriate time scaling $\epsilon_N$ should yield well-defined functions of order 1, while the fourth moment should be negligible. $\epsilon_N$ quantifies the scaling of the successive number of steps of the Markov chain that should be considered to obtain convergence to a diffusive process. 

\subsection{Application to simple population genetics models}
\paragraph{Wright-Fisher model.}
Consider the Wright-Fisher model with a fitness advantage $s$ of order $1/N$ of the mutant. Let us write $s=\alpha/N$ with $\alpha$ of order unity. The mean and variance of the variation of the mutant fraction $\Delta x$ read:
\begin{align}
	\mathbb{E}(\Delta x)&=\dfrac{\alpha x(1-x)}{N} +o\left(\frac{1}{N}\right)\,, \\
	\mathbb{V}(\Delta x)&= \dfrac{x(1-x)}{N}+o\left(\frac{1}{N}\right)\,.
\end{align}
We can check that the fourth moment satisfies $ \mathbb{E} \left(\Delta x^4\right)  = \mathcal{O}(1/N^2)$. Taking $\epsilon_N=1/N$, the assumptions of the theorem in the last section are satisfied, and thus $x_{[t/\epsilon_N]}$ converges weakly towards the diffusion process with generator
\begin{equation}
	\mathcal{L}_{\text{WF}}=\dfrac{1}{2}x(1-x) \dfrac{\partial^2}{\partial x^2} + \alpha x(1-x) \dfrac{\partial}{\partial x}\,.
\end{equation}

\paragraph{Moran model.}
For the Moran model,
\begin{align}
	\mathbb{E}(\Delta x)&=\dfrac{\alpha x(1-x)}{N^2} + o\left(\frac{1}{N^2}\right)\,, \\
	\mathbb{V}(\Delta x)&= \dfrac{2x(1-x)}{N^2}+o\left(\frac{1}{N^2}\right)\,,
\end{align}
and we can check that the fourth moment is $\mathcal{O}(1/N^3)$. This time, we define the time scaling as $\epsilon_N=1/N^2$, such that $x_{[t/\epsilon_N]}$ converges weakly to the diffusion process with generator
\begin{equation}
	\mathcal{L}_{\text{Moran}}=x(1-x) \dfrac{\partial^2}{\partial x^2} + \alpha x(1-x) \dfrac{\partial}{\partial x}.
\end{equation}
Note that we have some freedom in choosing the time scaling (up to a multiplicative constant), and that if another of these scalings was chosen, $x_{[t/\epsilon_N]}$ would simply converge to a different diffusion process with a rescaled generator. We choose this definition of $\epsilon_N$ to recover the usual diffusion equation for the Moran process. Other choices would just lead to a rescaling of time compared to it.

\subsection{Model with fixed updated fraction}

\paragraph{Model.} In our model with fixed updated fraction, we pick $M$ individuals and update them as we update the single individual in the Moran model (in our description of the Moran model presented in the main text), considering the state of the whole population. The fraction of updated individuals is $\rho=M/N$. Setting $\rho=1/N$ yields the Moran model, while setting $\rho=1$ gives the Wright-Fisher model. As explained in the main text, the probability of transition from $i$ mutants to $j$ mutants is
\begin{equation}
	P_{i\rightarrow j}^{(N,M)}=\sum_{k=0}^i \dfrac{\binom{i}{k}\binom{N-i}{M-k}}{\binom{N}{M}} \binom{M}{k+j-i} p^{k+j-i}(1-p)^{M+i-j-k}\,. \label{eq:transition_rho}
\end{equation}

\paragraph{Calculation of the first moments of $\Delta x$.}
Let us determine the first two moments of the variation $\Delta X$ of the number of mutants in the first discrete time step, given that we start with $i$ mutants. First, let us compute the mean of $\Delta X$:
\begin{align}
	\mathbb{E}(\Delta X) &=  \sum_{k= 0}^{\min(i,M)} \frac{\binom{i}{k} \binom{N-i}{M-k} }{\binom{N}{M}} \sum_{\Delta X} \Delta X \binom{M}{k+\Delta X} p^{k+\Delta X} ( 1 - p)^{M-\Delta X-k} = \sum_{k} \frac{\binom{i}{k} \binom{N-i}{M-k} }{\binom{N}{M}} \left(M p - k\right) \nonumber\\
	& = M p - \frac{1}{\binom{N}{M}} \sum_{k} \binom{i}{k} \binom{N-i}{M-k} k 
	= M p - i \frac{\binom{N-1}{M-1}}{\binom{N}{M}} = M p - \frac{Mi}{N}=M(p-x)\,,
\end{align}
where $x = i/N$.
Second, let us compute its second moment:
\begin{align}
	\mathbb{E}(\Delta X^2) &=   \sum_{k= 0}^{\min(i,M)}\frac{\binom{i}{k} \binom{N-i}{M-k} }{\binom{N}{M}} \sum_{\Delta X} \Delta X^2 \binom{M}{k+\Delta X} p^{k+\Delta X} ( 1 - p)^{M-\Delta X-k} \nonumber\\
	& = \sum_{k} \frac{\binom{i}{k} \binom{N-i}{M-k} }{\binom{N}{M}} \sum_{j} (j-k)^2 \binom{M}{j} p^{j} ( 1 - p)^{M-j} \nonumber\\
	& = Mp (1+p(M-1)) -\frac{ 2 M p}{\binom{N}{M}} \sum_{k} \binom{i}{k} \binom{N-i}{M-k} k +\frac{1}{\binom{N}{M}} \sum_{k} \binom{i}{k} \binom{N-i}{M-k} k^2 \nonumber\\& = Mp (1+p(M-1)) - 2  p \frac{iM^2}{N} + i (i-1) \frac{M(M-1)}{N(N-1)} + i \frac{M}{N}\,. \label{eq:deltaX2}
\end{align}
These results allow us to determine the mean and variance of $\Delta x= \Delta X/N$. We have
\begin{align}
	\mathbb{E}(\Delta x) =  \frac{\mathbb{E}(\Delta X)}{N} = \rho (p - x)\,,
\end{align}
where $\rho = M/N$. Using the expression of $p$ given in \cref{eq:xp} in the main text, and recalling that $s$ is assumed to be of order $1/N$, we obtain:
\begin{equation}
	p- x =\frac{x (1+ s)}{  1+ x s} - x = x \frac{ s - x s }{1+ x s} = s  \frac{x (1- x )}{1+ x s}  = s x (1- x ) + o \left(N^{-1}\right)\,,\label{eq:p-x}
\end{equation}
and hence 
\begin{align}
	\mathbb{E}(\Delta x) = \rho s x (1- x ) + o \left(\rho/N\right)\,.
\end{align}
In addition, we have
\begin{align}
	\mathbb{E}(\Delta x^2) &= \frac{\mathbb{E}(\Delta X^2)}{N^2} = \frac{1}{N^2} \left[ Mp (1+p(M-1)) - 2  p \frac{iM^2}{N} + i (i-1) \frac{M(M-1)}{N(N-1)} + i \frac{M}{N} \right]\nonumber\\
	& =  \rho p \left[\frac{1}{N}+p\left(\rho-\frac{1}{N}\right) \right] - 2 p x \rho^2 + x \rho \left(x-\frac{1}{N}\right) \frac{ \rho-N^{-1} }{1 -N^{-1}} +  \frac{\rho x}{N}\,.
\end{align}
Since $p = x + sx(1-x)+o(1/N)$ (see \cref{eq:p-x}), we find
\begin{equation}
	\mathbb{E}(\Delta x^2) =  \frac{\rho (2 - \rho)}{N} x(1 - x )  + o \left(\frac{\rho}{N}\right)\,,
	\label{eq:EDeltax2}
\end{equation}
and thus the variance is
\begin{equation}
	\mathbb{V}(\Delta x) =  \frac{\rho (2 - \rho)}{N} x(1 - x )  + o \left(\frac{\rho}{N}\right)\,. \label{eq:var_partial}
\end{equation}

\paragraph{Diffusion limit.} We can check that the  fourth moment of $\Delta x$ is $\mathcal{O}(\rho/N^2)$. 
Taking the scaling $\epsilon_N=\rho/N$, $x_{[t/\epsilon_N]}$  converges weakly towards the diffusion process with generator
\begin{equation}
	\mathcal{L}_{\rho}=\frac{2- \rho}{2}x(1-x) \dfrac{\partial^2}{\partial x^2} + \alpha  x(1-x) \dfrac{\partial}{\partial x}.
	\label{eq:generator_rho}
\end{equation}
Again, note that we have some freedom in choosing the time scaling (up to a multiplicative constant). We define $\epsilon_N$ this way to recover the usual Moran and Wright-Fisher diffusion equations in the limiting scenarios. Other choices would just lead to a rescaling of time compared to it.

\subsection{Fixation probability in the diffusion approximation}

\begin{figure}[h!]
	\includegraphics[width = 0.74\columnwidth]{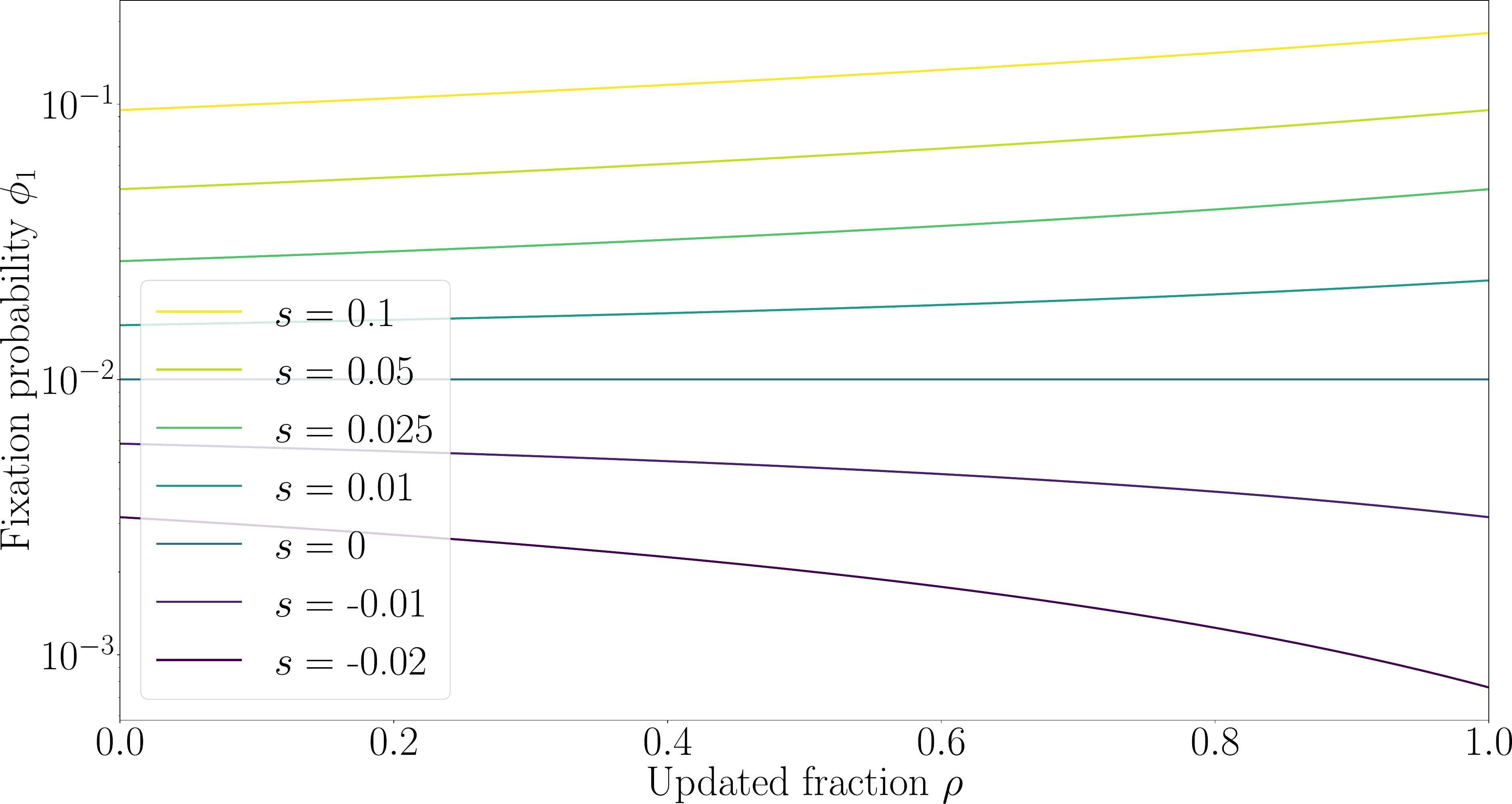}
	\caption{Fixation probability $\phi_1=\phi(1/N)$ (\cref{eq:fixproba} in the main text) as a function of the updated fraction $\rho$ for different relative fitness advantages $s$ of the mutant. Here $N= 100$ and the initial mutant fraction is $x = 1/N$.}
	\label{fig:phi_rho_different_s_x_1_N}
\end{figure}

\begin{figure}[h!]
	\includegraphics[width = 0.74\columnwidth]{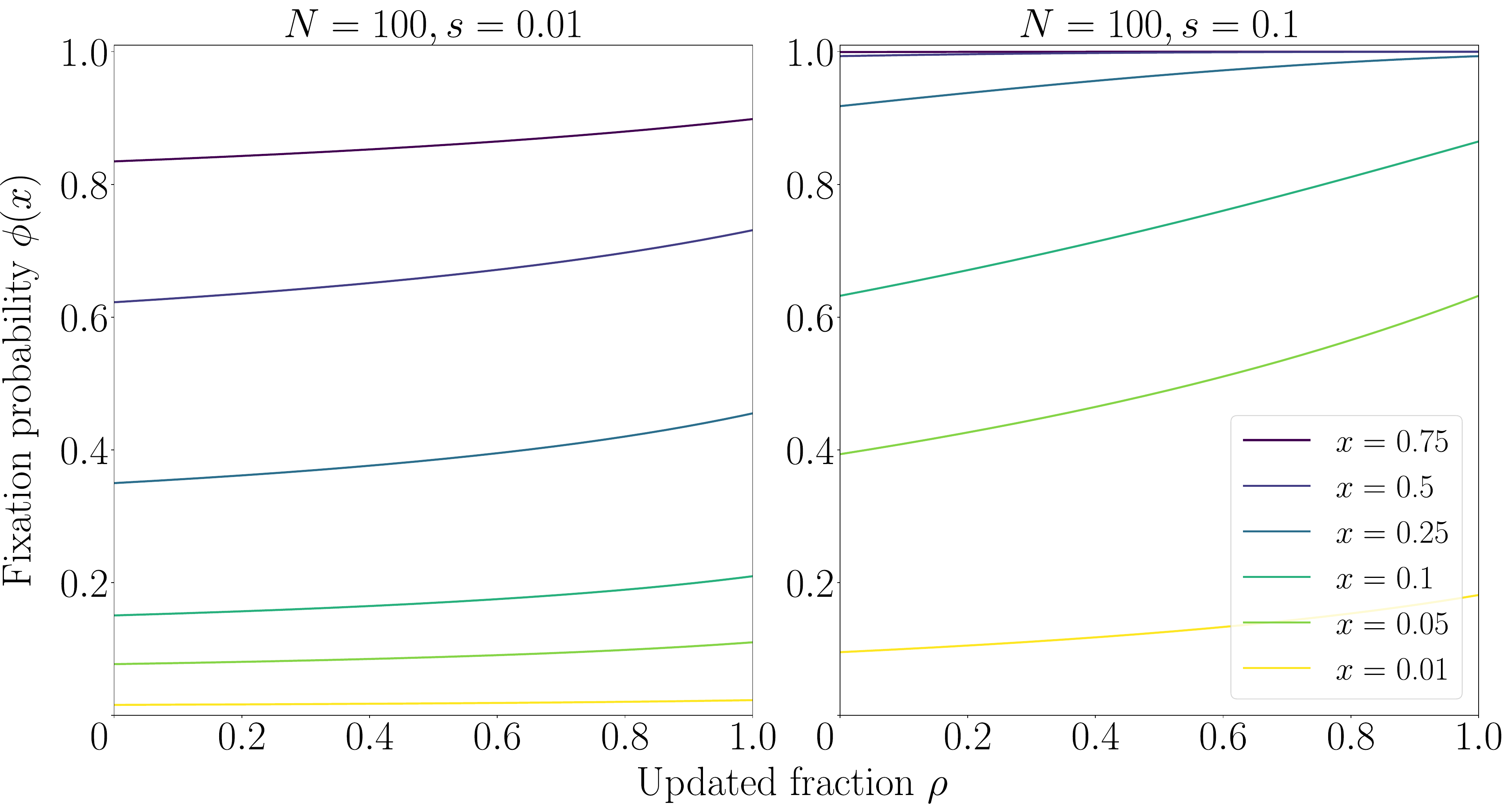}
	\caption{Fixation probability $\phi(x)$ (\cref{eq:fixproba} in the main text) as a function of the updated fraction $\rho$ for different initial mutant fractions $x$ with $N= 100$.}
	\label{fig:phisi2}
\end{figure}

Here, we provide additional plots of the mutant fixation probability in our partial update model, within the diffusion approximation. This fixation probability is given in Eq.~($11$) in the main text. \cref{fig:phi_rho_different_s_x_1_N,fig:phisi2} highlight the impact of the fraction $\rho$ of the population that is updated at each discrete time step on the mutant fixation probability. We observe that for most updated fractions $\rho$, our model behaves more similarly to a Moran model than to a Wright-Fisher model regarding fixation probabilities.

\newpage
\section{Mean absorption times}
In this section, we derive the expressions of the mean absorption times. We first consider the unconditional mean absorption time such that either one of the two absorbing barriers $x= 0$ and $x = 1$ (i.e.\ mutant extinction or fixation) is reached. Next, we condition the absorption on fixation and on extinction, leading to the mean fixation and extinction times.

\subsection{Mean unconditional absorption time}
We would like to determine the mean time $t(x)$ it takes to reach mutant extinction or fixation, starting from a given initial mutant fraction $x$. The unconditional mean absorption time $t(x)$, expressed in number of discrete time steps, satisfies the equation~\cite{ewens2004mathematical} 
\begin{equation}
	\mathcal{L}_{\rho} t(x) = -1\,,
\end{equation}
with the generator in \cref{eq:generator_rho}, and with the boundary conditions $t(0) = t(1)= 0$. The integration of this differential equation yields
\begin{equation}
	t(x) =  \frac{1-e^{-2ax/b}}{a(1-e^{-2a/b})} \int_{0}^1 \dd y \frac{1-e^{-2a(1-y)/b} }{y(1-y)} - \frac{1}{a}
	\int_{0}^{x} \dd y \frac{1-e^{-2a(x-y)/b} }{y(1-y)}\,,
\end{equation}
where $ a = \rho s$ and $b = \rho (2-\rho )/ N$.  By splitting the first integral in two, we can write
\begin{equation}
	t(x) =  \frac{1-e^{-2ax/b}}{a(1-e^{-2a/b})} \left\{\int_{0}^{x} \dd y  \frac{1-e^{-2a(1-y)/b} }{y(1-y)} +  \int_{x}^{1} \dd y \frac{1-e^{-2a(1-y)/b} }{y(1-y)} \right\} - \frac{1}{a}
	\int_{0}^{x} \dd y \frac{1-e^{-2a(x-y)/b} }{y(1-y)} \,.
\end{equation}
Rearranging terms yields 
\begin{equation}
	t(x) =  \frac{1-e^{2sN(1-x)/(2-\rho)}}{\rho s\left(1-e^{2sN/(2-\rho)}\right)} \int_{0}^{x} \dd y  \frac{e^{2sNy/(2-\rho)} -1}{y(1-y)} + \frac{1-e^{-2sNx/(2-\rho)}}{\rho s\left(1-e^{-2sN/(2-\rho)}\right)} \int_{x}^{1} \dd y  \frac{1-e^{-2sN(1-y)/(2-\rho)} }{y(1-y)}\,.
	\label{eq:uncond}
\end{equation}
This formula is consistent with the general solution $(4.21)$ given in Ref.~\cite{ewens2004mathematical}. 

\subsection{Mean conditional absorption times}
We can rewrite \cref{eq:uncond} as
\begin{equation}
	t(x) =   \int_{0}^{x} \dd y ~ t_1(x,y) + \int_{x}^{1} \dd y ~t_2(x,y)\,,
\end{equation}
where 
\begin{align}
	t_1(x,y)& = \frac{\left(1-e^{2sN(1-x)/(2-\rho)}\right)\left(e^{2sNy/(2-\rho)} -1\right)}{\rho sy(1-y)\left(1-e^{2sN/(2-\rho)}\right)}\,, \\
	t_2(x,y)& = \frac{\left(1-e^{-2sNx/(2-\rho)}\right)\left(1-e^{-2sN(1-y)/(2-\rho)}\right)}{\rho sy(1-y)\left(1-e^{-2sN/(2-\rho)}\right)}\,.
\end{align}

\paragraph{Mean mutant fixation time.} 
The mean absorption time conditioned on mutant fixation can then be expressed as~\cite{ewens2004mathematical}
\begin{equation}
	t^{*}(x) =   \int_{0}^{x} \dd y ~ t_1^{*}(x,y) + \int_{x}^{1} \dd y ~t_2^{*}(x,y)\,,
\end{equation}
with 
\begin{equation}
	t_{i}^{*}(x,y) = t_{i}(x,y) \frac{\phi(y)}{\phi(x)},~ \textrm{for}\,\,i\in\left\{1,2\right\}.
\end{equation}
where $\phi(x)$ is the mutant fixation probability starting from an initial mutant fraction $x$ given in \cref{eq:fixproba} in the main text. It yields
\begin{align}
	t^{*}(x)&=  \frac{1-e^{2sN(1-x)/(2-\rho)}}{\rho s(1-e^{2sN/(2-\rho)})(1-e^{-2sNx/(2-\rho)})} \int_{0}^{x} \dd y  \frac{\left(1-e^{-2sNy/(2-\rho)} \right) \left(e^{2sNy/(2-\rho)} -1\right)}{y(1-y)} \nonumber \\&+ \frac{1}{\rho s(1-e^{-2sN/(2-\rho)})} \int_{x}^{1} \dd y \frac{\left(1-e^{-2sN(1-y)/(2-\rho)}\right)\left(1-e^{-2sNy/(2-\rho)}\right) }{y(1-y)}\,,
	\label{eq:tfix}
\end{align}
which satisfies $t^{*}(1) =0$. 

In particular, when $s\ll 1$ and $Ns\gg1$, the leading term of the mean fixation time in \cref{eq:tfix} starting from one mutant ($x=1/N$) reads
\begin{equation}
	t^{*}\left(\frac{1}{N}\right)= \frac{2}{\rho s} \log\left(\frac{2Ns}{2-\rho}\right) .
\end{equation}
As this time is expressed in number of discrete time steps, in generations we have a mean fixation time
\begin{equation}
	\rho\, t^{*}\left(\frac{1}{N}\right)= \frac{2}{s} \log\left(\frac{2Ns}{2-\rho}\right) = \frac{2}{s} \log\left(2N_e s\right),
\end{equation}
where we introduced the effective population size $N_e=N/(2-\rho)$, see \cref{sec:effective_pop} of the main text. This expression is consistent with the fixation time of a beneficial mutant of order $(1/s) \log(Ns)$ generations found in Ref.~\cite{desai2007beneficial}, where stochastic birth and death events are modeled within a continuous-time branching process.\\

\paragraph{Mean mutant extinction time.} 
Similarly, the mean absorption time conditioned on mutant extinction is given by
\begin{equation}
	t^{**}(x) =   \int_{0}^{x} \dd y ~ t_1^{**}(x,y) + \int_{x}^{1} \dd y ~t_2^{**}(x,y)
\end{equation}
with 
\begin{equation}
	t_{i}^{**}(x,y) = t_{i}(x,y) \frac{\psi(y)}{\psi(x)},~\textrm{for}\,\, i\in\left\{1,2\right\}.
\end{equation}
where $\psi(x) = 1 - \phi(x)$ is the mutant extinction probability starting from an initial mutant fraction $x$. This gives
\begin{align}
	t^{**}(x)&=   \frac{1}{\rho s(1-e^{2sN/(2-\rho)})} \int_{0}^{x} \dd y \frac{\left(e^{2sNy/(2-\rho)} -1\right)\left(1-e^{2sN(1-y)/(2-\rho)}\right) }{y(1-y)} \nonumber\\&+
	\frac{1-e^{-2sNx/(2-\rho)}}{\rho s(1-e^{-2sN/(2-\rho)})(1-e^{2sN(1-x)/(2-\rho)})} \int_{x}^{1} \dd y  \frac{\left(1-e^{2sN(1-y)/(2-\rho)} \right) \left(1-e^{-2sN(1-y)/(2-\rho)} \right)}{y(1-y)} \,,
\end{align}
which satisfies $t^{**}(0) =0$. 

Note that the expressions obtained above satisfy the following symmetry property: $t^{*}(x) =t^{**}(1-x) $. To show this, let us consider 
\begin{align}
	t^{**}(1-x) &=   \frac{1}{\rho s(1-e^{2sN/(2-\rho)})} \int_{0}^{1-x} \dd y \frac{\left(e^{2sNy/(2-\rho)} -1\right)\left(1-e^{2sN(1-y)/(2-\rho)}\right) }{y(1-y)} \nonumber\\&+
	\frac{1-e^{-2sN(1-x)/(2-\rho)}}{\rho s(1-e^{-2sN/(2-\rho)})(1-e^{2sNx/(2-\rho)})} \int_{1-x}^{1} \dd y  \frac{\left(1-e^{2sN(1-y)/(2-\rho)} \right) \left(1-e^{-2sN(1-y)/(2-\rho)} \right)}{y(1-y)} 
\end{align}
A change of variable gives
\begin{align}
	t^{**}(1-x) &=   \frac{1}{\rho s(1-e^{2sN/(2-\rho)})} \int_{x}^{1} \dd y \frac{\left(e^{2sN(1-y)/(2-\rho)} -1\right)\left(1-e^{2sNy/(2-\rho)}\right) }{y(1-y)} \nonumber\\&+
	\frac{1-e^{-2sN(1-x)/(2-\rho)}}{\rho s(1-e^{-2sN/(2-\rho)})(1-e^{2sNx/(2-\rho)})} \int_{0}^{x} \dd y  \frac{\left(1-e^{2sNy/(2-\rho)} \right) \left(1-e^{-2sNy/(2-\rho)} \right)}{y(1-y)} \\&=   \frac{1}{\rho s(1-e^{-2sN/(2-\rho)})} \int_{x}^{1} \dd y \frac{\left(1-e^{-2sN(1-y)/(2-\rho)} \right)\left(1-e^{-2sNy/(2-\rho)}\right) }{y(1-y)} \nonumber\\&+
	\frac{1-e^{2sN(1-x)/(2-\rho)}}{\rho s(1-e^{2sN/(2-\rho)})(1-e^{-2sNx/(2-\rho)})} \int_{0}^{x} \dd y  \frac{\left(1-e^{-2sNy/(2-\rho)} \right) \left(e^{2sNy/(2-\rho)}-1 \right)}{y(1-y)} = t^{*}(x).
\end{align}
This symmetry is a consequence of the conditioning and of the fact that the operator $\mathcal{L}_{\rho}$ displays the symmetry $x \leftrightarrow 1-x$. Note however that in general $t(x) \neq t(1-x)$, due to the difference of fitness between  mutant and wild-type individuals. However, in the neutral case ($s=0$), we find 
\begin{align}
	t(x) = - \frac{2N}{\rho(2-\rho)} \left[ x\log(x) +(1-x)\log(1-x)\right]= - \frac{2N_e}{\rho} \left[ x\log(x) +(1-x)\log(1-x)\right],
\end{align}
and thus $t(x) = t(1-x)$ for $s =0$.

\section{Leading non-unit eigenvalue}

\subsection{Derivation from the Chia and Watterson model and link with the variance of viable offspring}

In Ref.~\cite{chia1969demographic}, Chia and Watterson presented a population genetics model for a population with a fixed size of $N$ individuals. In this model, each individual produces offspring according to a probability law represented by a branching process with generating function $f$. At each discrete step $t$, a number $m$ is drawn from a probability distribution $p_m$, and $m$ individuals are sampled from the offspring pool, yielding a total pool of individuals is of size $N+m$. Then, the number $r$ of survivors is drawn from a probability distribution $q_{mr}$, and $r$ individuals are picked from the parents and $N-r$ from the $m$ offspring. The distributions $p_m$ and $q_{mr}$ are chosen such that
$ 0 \leq r \leq N \text{ and } 0 \leq N-r \leq m$. For completeness, below we summarize results from~\cite{chia1969demographic}, and apply them to our partial update model.

Chia and Watterson computed all the eigenvalues of the transition matrix of the corresponding Markov process~\cite{chia1969demographic}. They also included a rate of mutation $\alpha_1$ from mutants to wild-types, and $\alpha_2$ for the opposite process.
In the neutral case, the transition matrix element from $i$ to $j$ mutants in this model reads
\begin{multline}
	P_{i\rightarrow j}=\sum_{m=0}^\infty \sum_{b=0}^\infty \sum_{r=0}^\infty p_m c_{mN} q_{mr} \text{ coefficient of }w^m z_1^b z^j \text{ in } \\ H(i,N-i,r,z)H(b,m-b,N-r,z) f^{i}((1-\alpha_1)z_1 w +\alpha_1 w) f^{N-i}(\alpha_2 z_1 w + (1-\alpha_2)w),  \label{eq:pij_theory}
\end{multline}
where
\begin{equation}
	H(i, N-i, r, z) =\sum_{k=0}^\infty z^k \dfrac{\binom{i}{k} \binom{N-i}{r-k}}{\binom{N}{r}},
\end{equation}
and $c_{mN}$ is the inverse of the coefficient of $z^m$ in $f^N(z)$. Focusing on the coefficient of $w^m$ means that we constrain the total offspring number to $m$. Focusing on the coefficient of $z_1^b$ means we are looking at cases in which the number of mutants within the offspring is $b$. Finally, focusing on the coefficient of $z^j$ means that the total number of mutants (among the population once it is updated) is $j$. The function $H(i, N-i, r, z)$ represents the ways of sampling $r$ individuals from a population of $N$ parents comprising $i$ mutants and $N-i$ wild-types. The function $H(b, m-b, N-r, z)$ represents the ways of sampling $N-r$ individuals from the $m$ offspring population made of $b$ mutants and $m-b$ wild-types. The function $f^{i}((1-\alpha_1)z_1 w +\alpha_1 w)$ is the generating function for $i$ mutants, and $f^{N-i}(\alpha_2 z_1 w + (1-\alpha_2)w)$ is the generating function for $N-i$ wild-types. \\

\paragraph{Application: partial update with fixed number of offspring and survivors.} Our partial update model can be seen as a particular case of the Chia and Watterson model, where we produce exactly the same number of offspring $M$ at each step, keep those $M$ individuals, and complete the new generation with $N-M$ individuals from the previous generation. This means that $p_m=1$ if $m=M$ and $0$ otherwise. We also have $q_{mr}=1$ if $r=N-M$, and 0 otherwise. We consider the no-mutation case where $\alpha_1 = \alpha_2=0$. For the branching process representing offspring production, we use the Poisson law $\mathscr{P}(\mu)$ with generating function $f(z)=\exp(\mu(z-1))$. Let us write the transition matrix elements as $p_{ij}$, representing the probability of going from $i$ to $j$ mutants. We have
\begin{align}
	f^i(z_1 w) f^{N-i}(w)&=e^{-\mu N} \sum_k \dfrac{\mu^k}{k!}(i z_1 w)^k \sum_l \dfrac{\mu^l}{l!}(N-i)^l w^l ,\\
	f^N(w)&=e^{-\mu N} \sum_k \dfrac{\mu^k}{k!}w^k N^k.
\end{align}
Computing the coefficient of $w^m z_1^b z^j$ in the expression given in \cref{eq:pij_theory} yields
\begin{align}
	\text{coef} = e^{-\mu N} \sum_{k = 0}^{i} \frac{\binom{i}{k} \binom{N-i}{r-k}}{\binom{N}{r}} \frac{\binom{b}{j-k} \binom{m-b}{N-r-j+k}}{\binom{m}{N-r}} \frac{(i\mu)^b}{b!}\frac{\mu^{m-b}(N-i)^{m-b}}{(m-b)!},
\end{align}
and we have
\begin{equation}
	c_{mN} = \frac{m!\, e^{\mu N}}{(\mu N)^m}.
\end{equation}

Putting everything together with \cref{eq:pij_theory} yields
\begin{equation}
	P_{i\rightarrow j}^{(N,M)}=\sum_{b=0}^{\infty} \sum_{k=0}^{i}\dfrac{\binom{i}{k} \binom{N-i}{N-M-k}}{\binom{N}{N-M}} \dfrac{\binom{b}{j-k} \binom{M-b}{M-j+k}}{\binom{M}{M}} \binom{M}{b} \left(\dfrac{i}{N} \right)^{b} \left(1-\dfrac{i}{N} \right)^{M-b}.
\end{equation}
Note that the parameter $\mu$ cancels out in this expression. Noticing that $\binom{b}{j-k} \binom{M-b}{M-j+k}$ is $1$ only if $b = j-k$ and $0$ otherwise, the equation simplifies, and we ultimately find  
\begin{equation}
	P_{i\rightarrow j}^{(N,M)}=\sum_{k=0}^{i} \dfrac{\binom{i}{k} \binom{N-i}{M-k}}{\binom{N}{M}} \binom{M}{k+j-i} \left(\dfrac{i}{N} \right)^{k+j-i} \left(1-\dfrac{i}{N} \right)^{M-k-j+i}, \label{eq:Pij_neutral}
\end{equation}
which is the same as \cref{eq:transition_rho} in the neutral case.\\

\paragraph{Computing the relevant eigenvalue.}
Chia and Watterson provided an explicit expression of all eigenvalues of the transition matrix for neutral mutants in Ref.~\cite{chia1969demographic}. Here, we are interested in the second largest eigenvalue $\lambda_*$ (the leading eigenvalue is equal to 1 and has multiplicity 2 when there are no mutations). Applying Chia and Watterson's formula to our partial update model yields
\begin{align}
	\lambda_* &= c_{MN} \binom{2}{0} \dfrac{(N-M)(N-M-1)}{N(N-1)} \text{ coeff. of } w^M \text{ in } f^N(w) \nonumber\\ &+ c_{MN} \binom{2}{1} \dfrac{N-M}{N} \text{ coeff. of } w^{M-1} \text{ in } f'(w)f^{N-1}(w) \nonumber\\&+ c_{MN} \binom{2}{2} \text{ coeff. of } w^{M-2} \text{ in } [f'(w)]^2 f^{N-2}(w).
\end{align}
With a little algebra, we reach
\begin{equation}
	\lambda_*= \dfrac{(N-M)(N-M-1)}{N(N-1)} + 2 \dfrac{(N-M)M}{N^2}+\dfrac{M(M-1)}{N^2},
\end{equation}
and substituting $M/N$ with $\rho$, we finally obtain:
\begin{equation}
	\lambda_*=1-\dfrac{\rho(2-\rho)}{N} + o(\rho/N). \label{eq:lambda2}
\end{equation}

\paragraph{Link with the variance of viable offspring.}
Chia and Watterson define \emph{viable} offspring as individuals that are part of the $m$ offspring at one generation, and that are kept to be part of the next generation, i.e.\ that are among the $N-r$ individuals  sampled from the offspring.  A viable individual has a life time which is geometrically distributed, and the mean lifetime is
\begin{equation}
	\mu = \left[ \sum_m \sum_r p_m q_{mr} \dfrac{N-r}{N} \right]^{-1}. \label{eq:lifetime}
\end{equation}
Looking at the survival of one viable individual across several generations, we can write the generating function of its number of viable offspring $V$, $\phi(z)=\mathbb{E}[z^V]$. The detailed definition of this function is given in Ref.~\cite{chia1969demographic}. From this expression, we can derive that $\phi'(1)=1$, which means that the average number of viable offspring produced by one individual is one, consistent with the fact that the population size remains constant. We can also derive the variance $\phi''(1)$ which appears in the eigenvalue $\lambda_*$ (see Ref.~\cite{chia1969demographic}) as
\begin{equation}
	\lambda_* = 1 - \dfrac{\phi''(1)}{\mu (N-1)}. \label{eq:lambda_star_gen}
\end{equation}
For our partial update model, the sums over $m$ and $r$ in the mean lifetime $\mu$ simplify and we get
\begin{equation}
	\mu= \frac{N}{M} = \frac{1}{\rho}.
\end{equation}
From \cref{eq:lambda_star_gen}, and the expression of $\lambda_*$ in \cref{eq:lambda2}, we obtain for our partial update model $\phi''(1)=2-\rho$. From the diffusion approximation applied to the partial update model, and assuming $s\ll 1$ while $Ns\gg 1$ and $N\gg 1$, the fixation probability of one individual can be written as
\begin{equation}
	\phi_1=\dfrac{2s}{2-\rho}.
\end{equation}
This is consistent with Haldane's expression of the fixation probability which reads $\phi_1=2s/\sigma^2$, where $\sigma^2$ is the variance of the offspring, and thus is equal to $\phi''(1)=2-\rho$.

\subsection{Derivation of the leading non-unit eigenvalue using Ref.~\cite{ewens2004mathematical}}
Here, we present another method proposed in~\cite{ewens2004mathematical} to compute the leading non-unit eigenvalue $\lambda_*$ of the transition matrix for the partial update model in the neutral case. The key result is that, if there exists a function $f$ such that $\mathbb{E}[f(X_{t+1})|X_t]=a f(X_t)$ where $a$ is a constant, then $a=\lambda_*$ is the desired eigenvalue~\cite{ewens2004mathematical}. Taking $f(X)=X(N-X)$ for our partial update model yields
\begin{align}
	\mathbb{E}(X_{t+1}|X_t)  &= X_t \, \\
	\mathbb{E}\left[X_{t+1}(N-X_{t+1})|X_t\right] & = \mathbb{E}\left[(X_{t}+ \Delta X)(N-X_{t}- \Delta X)|X_t\right] \nonumber\\
	& = X_t(N-X_t) - \mathbb{E}(\Delta X^2|X_t)  \nonumber \\
	& = X_t(N-X_t)\left[1- \frac{\rho(2-\rho)}{N} + o\left(\frac{\rho}{N}\right) \right] ,
\end{align}
where we used the fact that $\mathbb{E}(\Delta X)=0$ in the neutral case, and the expression of $\mathbb{E}(\Delta X^2)$ from \cref{eq:deltaX2,eq:EDeltax2}. Therefore, we can identify
\begin{equation}
	\lambda_{*} = 1- \frac{\rho(2-\rho)}{N} + o\left(\frac{\rho}{N}\right),
\end{equation}
which matches the result obtained in \cref{eq:lambda2} using the Chia and Watterson model.

\subsection{Rate of approach to the stationary state}

The largest non-unit eigenvalue $\lambda_{*}$ of the transition matrix $\boldsymbol{P} = \left\{P_{i\rightarrow j} \right\}_{ij}$ quantifies the rate of convergence towards the stationary state. Concretely, let us consider 
the vector $\boldsymbol{\pi}_{t} = (p_0(t),p_1(t),p_2(t),\dots , p_N(t))$ where $p_k(t)$ denotes the probability of having $k$ mutants at time $t$ starting from $1$ mutant at $t =0$. This vector satisfies
\begin{equation}
	\boldsymbol{\pi}_{t+1} = \boldsymbol{P} \,\boldsymbol{\pi}_{t}\,,
\end{equation}
with $\boldsymbol{\pi}_{0} = (0,1,0,0,\dots ,0)$ at time $t=0$. We can decompose this vector in the basis of eigenvectors of $\boldsymbol{P}$ written $\boldsymbol{u}_k$:
\begin{equation}
	\boldsymbol{\pi}_{0} =  
	\sum_{k = 0}^{N} a_k \boldsymbol{u}_k\,,
\end{equation}
where we introduced coefficients $a_k$. At time step $t$, we have
\begin{equation}
	\boldsymbol{\pi}_{t} = {\boldsymbol{P}}^t \boldsymbol{\pi}_{0} =  
	\sum_{k = 0}^{N} a_k {\lambda_k}^t    \boldsymbol{u}_k\,,
\end{equation}
where $\lambda_0 = \lambda_1 = 1$, due to the two absorbing states (extinction and fixation of the mutants) and $0<\lambda_k < 1$ for $k\geq 2$. When $t\rightarrow+\infty$, we obtain
\begin{equation}
	\lim\limits_{t\rightarrow+\infty}\boldsymbol{\pi}_{t} =\boldsymbol{\pi}_{\infty} =  a_0\boldsymbol{u}_0
	+a_1\boldsymbol{u}_1\,,
\end{equation}
where $\boldsymbol{\pi}_{\infty} = (1-1/N,0,0,\dots ,0,1/N)$.
We can thus write
\begin{equation}
	\boldsymbol{\pi}_{t} =  \boldsymbol{\pi}_{\infty} +a_2 {\lambda_{*}}^t \boldsymbol{u}_2 +  
	\sum_{k = 3}^{N} a_k {\lambda_k}^t \boldsymbol{u}_k
\end{equation}
By taking the Euclidean norm $||\boldsymbol{\pi}_t - \boldsymbol{\pi}_{\infty} ||_{2}$ of the difference between $\boldsymbol{\pi}_{t}$ and $\boldsymbol{\pi}_{\infty}$, we find
\begin{equation}
	||\boldsymbol{\pi}_t - \boldsymbol{\pi}_{\infty} ||_{2} =  {\lambda_{*}}^t a_2 ||\boldsymbol{u}_2||_{2} +  
	\sum_{k = 3}^{N}  {\lambda_k}^t a_k ||\boldsymbol{u}_k||_{2}\,.
	\label{eq:convergence_markov}
\end{equation}
In the long-time limit, the dominant term is the one involving $\lambda_{*}$. Thus, for $t\rightarrow \infty$,
\begin{equation}
	||\boldsymbol{\pi}_t - \boldsymbol{\pi}_{\infty} ||_{2}\propto {\lambda_{*}}^t = {r}^{\rho t} \,,
\end{equation}
where $ r = {\lambda_{*}}^{1/\rho}$ (see \cref{eq:rate} in the main text).

\section{Coalescent process}

In this section, we study the coalescent process associated with our model with partial updates, in the case of neutral mutants. Importantly, in this specific case, our model falls in the broad class of Cannings models~\cite{cannings1974latent}. Since these models converge to Kingman's coalescent~\cite{kingman1982a, kingman1982b} in the limit of large population sizes and under the proper time scaling~\cite{mohle1998,mohle2001classification}, our model does too. For completeness, we expand on this point, and we present explicit calculations for finite population size, yielding in particular an explicit expression of the coalescent effective population size in our model.

\subsection{Convergence to Kingman's coalescent and associated time scaling}
\label{sec:Kingman}

In Refs.~\cite{kingman1982a, kingman1982b}, Kingman studied the genealogy of haploid populations with large fixed size $N$, and introduced the limiting ancestral process for a broad class of population models. The idea is to trace the history of a sample group of $n\ll N$ lineages, which go through coalescence events. Each coalescence event decreases the number of lineages by 1. Kingman first analyzed the Wright-Fisher model with population size $N$. Let us denote by $G_{i,j}$ the probability that $i\leq n$ individuals, sampled at time $t$ among the $n$ individuals considered, have exactly $j$ ancestors at time $t-1$, i.e.\ at the previous generation. Kingman proved that the $G_{i,j}$ converge to the following process in the large-$N$ limit:
\begin{align}
	G_{i,i} &= 1 - \dfrac{\binom{i}{2}}{N} + o\left(\frac{1}{N}\right) \quad \text{for }i=1,\dots ,n. \label{eq:GiiK}\\
	G_{i,i-1}&= \dfrac{\binom{i}{2}}{N} + o\left(\frac{1}{N}\right) \quad \text{for }i=1,\dots ,n. \label{eq:Gii-1K} \\
	G_{i,j}&= o\left(\frac{1}{N}\right) \quad \text{for }i=1,\dots ,n \text{ and }j\neq i,\,\, j\neq i-1. \label{GijK}
\end{align}
Note that the assumption $n\ll N$ and the fact that $i\leq n$ jointly ensure that $\binom{i}{2} = \mathcal{O}(1)$. 
Kingman also underlined the robustness of this result. Consider a model for a population of size $N$, where the offspring number of each individual $i$ is a random variable $\nu_i$. Further assume that the $\nu_i$ are sampled from an exchangeable joint distribution. This assumption holds for the neutral case in our partial update model, since all individuals have the same offspring number probability distribution. In particular, we have $\mathbb{E}[\nu_1]=1$, as each individual has on average one offspring in this model where the population size is fixed. Assume that as $N \rightarrow \infty$, the variance of $\nu_1$ goes to a finite non-zero limit $v$ and that its moments are bounded by some values $M_k$:
\begin{align}
	&\mathbb{V}[\nu_1] \rightarrow v\,, \\
	&\mathbb{E}[\nu_1^k] \leq M_k \quad \text{for }k \geq 2.
\end{align}
Kingman stated that under these conditions, the scaled continuous time process converges to a process called the \emph{$n$-coalescent}, also known as Kingman's coalescent (note that minimal assumptions to ensure convergence to Kingman's coalescent were found in~\cite{sagitov2003convergence, mohle2003coalescent}, and only regard the second and third moments of $\nu_1$, i.e.\ $k=2$ and $k=3$ here). The proper time scaling to obtain this convergence is to look at the discrete process every $N/v$ steps. This universality was further justified by Möhle~\cite{mohle1998}, for the general setting of the Cannings model~\cite{cannings1974latent}, which includes our partial update model in the specific case of neutral mutants. 

Hence, the coalescent process associated with our partial update model converges to Kingman's coalescent for large population sizes. Specifically, considering neutral mutants in our model, we can express $\mathbb{V}[\nu_1]$ by considering the case where only one neutral mutant is initially present and calculating the variance of the variation of mutant number in one step. Thus, using \cref{eq:var_partial} with $x=1/N$ allows us to reach
\begin{equation}
	\mathbb{V}[\nu_1]\rightarrow \rho(2-\rho).
	\label{eq:Vnu1}
\end{equation}
This entails that, as long as $\rho$ does not tend to zero when $N\rightarrow \infty$, the corresponding rescaled continuous process converges to Kingman's coalescent. In addition, the proper time scaling is to consider the discrete process every $c_N = N/[\rho (2 - \rho)]$ time steps. 
Note that this time scaling is different from the one we chose above to obtain convergence to the diffusion process. The latter considered the discrete process taken every $1/\epsilon_N=N/\rho$ discrete time steps. While $\epsilon_N$ could be defined up to a multiplicative constant to ensure convergence, the time scaling for the coalescent has a unique definition, see also~\cite{mohle2001classification}.

Note that models where $\rho\to 0$ when $N\to \infty$, and in particular the Moran model, are degenerate cases, with $\mathbb{V}[\nu_1] \rightarrow 0$. They should thus be treated separately. However, it was shown that the Moran model yields the same result, as its ancestral process also converges to Kingman's coalescent with the proper scaling of $N^2/2$ time steps (which corresponds to $c_N$ with $\rho=1/N$). Studies of the coalescent process for the Wright-Fisher and Moran models can be found in~\cite{watterson1975,hudson1983,tajima1983}, and are reviewed in~\cite{nordborg2019, donnelly1995coalescents}.

\subsection{Coalescent process for finite population size}
As stated in the previous section, the ancestral process of our partial update model converges to Kingman's coalescent in the limit $N \rightarrow \infty$, under the proper time scaling~\cite{kingman1982a, kingman1982b}. For the sake of completeness, let us now derive the general expressions of the $G_{i,j}$ for finite $N$. Next, in \cref{sec:Gladstien}, we will show that these expressions are consistent with general results known for the Chia and Watterson model.

To study the coalescent, let us consider our model with fixed updated size $M$
in the neutral case ($s=0$). We would like to determine explicitly the probability $G_{i,j}$ that $i$ sampled individuals at time $t$ have exactly $j$ ancestors at time $t-1$, for finite $N$. For this, let us focus on one discrete time step. Let us first compute the probability $G_{i,i}$ that no coalescence occurs: 
\begin{align}
	G_{i,i} &= \frac{\binom{N-M}{i}\binom{M}{0}}{\binom{N}{i}}+ \frac{\binom{N-M}{i-1}\binom{M}{1}}{\binom{N}{i}} \left(1-\frac{i-1}{N}\right)  \nonumber \\ &+ \frac{\binom{N-M}{i-2}\binom{M}{2}}{\binom{N}{i}} \left(1-\frac{i-2}{N}\right)\left(1-\frac{i-1}{N}\right) \nonumber \\ &+ \frac{\binom{N-M}{i-3}\binom{M}{3}}{\binom{N}{i}}\left(1-\frac{i-3}{N}\right) \left(1-\frac{i-2}{N}\right)\left(1-\frac{i-1}{N}\right) \nonumber  \\ &+ \dots + \frac{\binom{N-M}{i-k}\binom{M}{k}}{\binom{N}{i}}\left(1-\frac{i-k}{N}\right) \dots \left(1-\frac{i-2}{N}\right)\left(1-\frac{i-1}{N}\right)+ \dots 
	\nonumber\\ & = \sum_{k=0}^{\min(i, M)} \frac{\binom{N-M}{i-k}\binom{M}{k}}{\binom{N}{i}} \frac{(N-i+k)!}{(N-i)!N^k}. \label{eq:Gii_gen}
\end{align}
In this equation, the first term corresponds to the probability that none of the $i$ sampled individuals are newly born (i.e. their ancestor was not among the $M$ updated individuals in the population): then, they necessarily have $i$ ancestors at $t-1$, themselves. The second term corresponds to the probability that 1 among the $i$ sampled individuals is newly born, and that its lineage does not coalesce with any of the lineages of the $i-1$ other sampled individuals. The third term corresponds to the probability that 2 among the $i$ sampled individuals are newly born, and that the lineage of the first one does not coalesce with any of the lineages of the $i-2$ other sampled individuals that are not newly born (thus increasing the pool of ancestor to $i-1$), and that the lineage of the second one does not coalesce with the lineages of the $i-1$ other sampled individuals, etc. 

In addition, we have
\begin{align}
	G_{i,i-1} &=  \frac{\binom{N-M}{i-1}\binom{M}{1}}{\binom{N}{i}} \frac{i-1}{N} \nonumber \\ &+ \frac{\binom{N-M}{i-2}\binom{M}{2}}{\binom{N}{i}} \left[ \frac{i-2}{N}\left(1-\frac{i-2}{N}\right) +\left(1-\frac{i-2}{N}\right) \frac{i-1}{N}\right] \nonumber \\ &+ \frac{\binom{N-M}{i-3}\binom{M}{3}}{\binom{N}{i}} \left[\frac{i-3}{N}\left(1-\frac{i-3}{N}\right)\left(1-\frac{i-2}{N}\right) +\left(1-\frac{i-3}{N}\right)\frac{i-2}{N} \left(1-\frac{i-2}{N}\right) \right. \nonumber \\& \left. + \left(1-\frac{i-3}{N}\right)\left(1-\frac{i-2}{N}\right)\frac{i-1}{N}  \right]\nonumber  \\ &+ \dots + \frac{\binom{N-M}{i-k}\binom{M}{k}}{\binom{N}{i}}\left(1-\frac{i-k}{N}\right) \dots \left(1-\frac{i-2}{N}\right)\sum_{\ell = 1}^{k} \frac{i-\ell}{N} + \dots  \nonumber
	\\ & = \sum_{k=1}^{\min(i, M)} \frac{\binom{N-M}{i-k}\binom{M}{k}}{\binom{N}{i}} \frac{(N-i+k)!}{(N-i+1)!N^k} \sum_{\ell_1 = i-k}^{i-1}\ell_1.
	\label{eq:Gii-1_gen}
\end{align}
In this equation, the first term corresponds to the probability that 1 among the $i$ sampled individuals is newly born, and that its lineage coalesces with one of the other $i-1$ lineages. The next term corresponds to the probability that 2 among the $i$ sampled individuals (that we will call A and B) are newly born, and then describes two possibilities. The first one is the following: the lineage of A coalesces with one of the $i-2$ other (non-A and non-B) lineages, leaving $i-2$ individuals in the current pool of ancestors, whereas the lineage of B does not coalesce with any of these $i-2$ lineages, thus increasing the pool of ancestors to $i-1$. The second possibility is the following: the lineage of A does not coalesce with one of the $i-2$ other lineages, increasing the pool of ancestors to $i-1$, whereas the lineage of B coalesces with one of these $i-1$ lineages, leaving a pool of ancestors of size $i-1$. The other terms can be constructed in a similar way.

We can also obtain the next term
\begin{align}
	G_{i,i-2} &=  \frac{\binom{N-M}{i-2}\binom{M}{2}}{\binom{N}{i}} \frac{i-2}{N}\frac{i-2}{N} \nonumber \\ &+ \frac{\binom{N-M}{i-3}\binom{M}{3}}{\binom{N}{i}} \left[\frac{i-3}{N}\frac{i-3}{N}\left(1-\frac{i-3}{N}\right) +\left(1-\frac{i-3}{N}\right)\frac{i-2}{N} \frac{i-2}{N} + \frac{i-3}{N}\left(1-\frac{i-3}{N}\right)\frac{i-2}{N}  \right]\nonumber  \\ &+ \frac{\binom{N-M}{i-4}\binom{M}{4}}{\binom{N}{i}} \left[\frac{i-4}{N}\frac{i-4}{N}\left(1-\frac{i-4}{N}\right)\left(1-\frac{i-3}{N}\right) + \frac{i-4}{N}\left(1-\frac{i-4}{N}\right) \frac{i-3}{N} \left(1-\frac{i-3}{N}\right) \right. \nonumber\\ &+\frac{i-4}{N}\left(1-\frac{i-4}{N}\right) \left(1-\frac{i-3}{N}\right)  \frac{i-2}{N} + \left(1-\frac{i-4}{N}\right)\frac{i-3}{N} \left(1-\frac{i-3}{N}\right)  \frac{i-2}{N}\nonumber \\ &\left.+\left(1-\frac{i-4}{N}\right)  \left(1-\frac{i-3}{N}\right)  \frac{i-2}{N} \frac{i-2}{N} + \left(1-\frac{i-4}{N}\right)\frac{i-3}{N} \frac{i-3}{N} \left(1-\frac{i-3}{N}\right) \right] \nonumber \\ &+ \dots +\frac{\binom{N-M}{i-k}\binom{M}{k}}{\binom{N}{i}}\left(1-\frac{i-k}{N}\right) \dots \left(1-\frac{i-3}{N}\right)\sum_{\ell_1 = i-k}^{i-2}\sum_{\ell_2 = \ell_1}^{i-2}\frac{\ell_1\ell_2}{N^2} + \dots  \nonumber
	\\ & = \sum_{k=2}^{\min(i, M)} \frac{\binom{N-M}{i-k}\binom{M}{k}}{\binom{N}{i}} \frac{(N-i+k)!}{(N-i+2)!N^k} \sum_{\ell_1 = i-k}^{i-2}\sum_{\ell_2 = \ell_1}^{i-2}\ell_1 \ell_2.
\end{align}

These formulas can be extrapolated to express the general case:
\begin{align}
	G_{i,j} = \sum_{k=i-j}^{\min(i, M)} \frac{\binom{N-M}{i-k}\binom{M}{k}}{\binom{N}{i}} \frac{(N-i+k)!}{(N-j)!N^k} {i\brace j}_ {i-k}\,,
	\label{eq:Gijgal}
\end{align}
with the $r$-Stirling number of the second kind defined as~\cite{broder1984r}
\begin{align}
	{i\brace j}_ {r} =  \sum_{\ell_1 = r}^{j}\sum_{\ell_2 = \ell_1}^{j} \dots  \sum_{\ell_{i-j} = \ell_{i-j-1}}^{j}\ell_1 \ell_2 \dots \ell_{i-j}\,,
\end{align}
which satisfies 
\begin{align}
	{i\brace j}_ {r} = 0&\hspace{1cm}\text{if } i<r, \\
	{i\brace j}_ {r}  = \delta_{j,r}&\hspace{1cm}\text{if } i=r,\\
	{i\brace j}_ {r} = j {i-1\brace j}_ {r} + {i-1\brace j-1}_ {r}&\hspace{1cm}\text{if } i>r,
\end{align}
where $\delta_{j,r}$ is the Kronecker symbol, which is 1 if $j=r$ and 0 otherwise.

\subsection{Link with the eigenvalues of the transition matrix}
\label{sec:Gladstien}
In Ref.~\cite{gladstien1978}, considering a general framework of constant-size haploid models without selection, which includes the Chia and Watterson model, Gladstien showed that $G_{i,i}=\lambda_i$ for $i=1,\dots ,N$, where $\lambda_0, \cdots, \lambda_N$ are the $N+1$ eigenvalues of the transition matrix of the discrete process. The elements of the transition matrix in our model in the neutral case are given in \cref{eq:Pij_neutral}. We already discussed the two largest eigenvalues $\lambda_0=\lambda_1=1$, and the third largest eigenvalue $\lambda_2 = \lambda_*$ in \cref{eq:lambda_star_gen}. The general expression of the eigenvalues associated with our model in the neutral case can be obtained as a particular case of those derived by Chia and Watterson~\cite{chia1969demographic}, and read for all $i$:
\begin{align}
	\lambda_i &=  \sum_{k=0}^{\min(i, M)} \binom{i}{k} \dfrac{(N-M)!}{(N-M-i+k)!} \dfrac{M!}{(M-k)!} \dfrac{(N-i+k)!}{N!}\dfrac{1}{N^k} \nonumber\\
	&= \sum_{k=0}^{\min(i, M)} \frac{\binom{N-M}{i-k}\binom{M}{k}}{\binom{N}{i}} \frac{(N-i+k)!}{(N-i)!N^k}.
\end{align}
We recover \cref{eq:Gii_gen}, as expected. Besides, Gladstien~\cite{gladstien1978} and Felsenstein~\cite{felsenstein1971rate} state that, for $i=2,\dots ,N$:
\begin{equation}
	G_{i-1,i-1}=G_{i,i}+\dfrac{2}{i} \,G_{i,i-1}, \label{eq:gladstien}
\end{equation}
from which we recover the expression of $G_{i,i-1}$ in \cref{eq:Gii-1_gen}.

\subsection{Large-$N$ limit}

To make the link between the finite-$N$ results above and Kingman's coalescent, let us consider the large-$N$ limits of the finite-$N$ results.\\

\paragraph{Large-$N$ limit of $G_{i,i}$.}
We can rewrite the probability $G_{i,i}$ of having no coalescence event in one discrete time step in the limit $N\rightarrow \infty$ and $i \ll N$. 
\begin{equation}
	G_{i,i} = \sum_{k=0}^{\min(i, M)} \frac{\binom{N-M}{i-k}\binom{M}{k}}{\binom{N}{i}} \frac{(N-i+k)!}{(N-i)!N^k} =\sum_{k=0}^{\min(i, M)} q_{N,i,M}(k) \prod_{j=1}^k \left( 1 - \dfrac{i-j}{N}\right)\,, \label{eq:Gii_hyper}
\end{equation}
where $q_{N,i,M}(k)$ is the probability of sampling $k$ from a hypergeometric distribution with parameters $N, i, M$.
Let us expand the product 
\begin{equation}
	\prod_{j=1}^k \left( 1 - \dfrac{i-j}{N}\right)=1 - \sum_{j=1}^k \dfrac{i-j}{N}+\dfrac{1}{2}\sum_{\substack{j_1, j_2=1 \\ j_1 \neq j_2}}^k\dfrac{(i-j_1)(i-j_2)}{N^2}+\dfrac{1}{3!} \sum_{\substack{j_1, j_2, j_3=1 \\ j_1, j_2, j_3\,\, \textrm{p.d.}}}^k\dfrac{(i-j_1)(i-j_2)(i-j_3)}{N^3}
	+ \dots + \prod_{j=1}^k\dfrac{i-j}{N},
	\label{eq:expand}
\end{equation}
where the notation ``p.d.'' means pairwise distinct. Let us consider the first two terms of \cref{eq:Gii_hyper} coming from the expansion in \cref{eq:expand}:
\begin{equation}
	\sum_{k=0}^{\min(i, M)} q_{N,i,M}(k)   \left[ 1 - \sum_{j=1}^k \dfrac{i-j}{N}\right] = \sum_{k=0}^{\min(i, M)} q_{N,i,M}(k)   \left[ 1 - \dfrac{ki}{N}+\dfrac{k(k-1)}{2N}\right]= 1 - \dfrac{i+\frac{1}{2}}{N} \mathbb{E}_q[k]+\dfrac{1}{2N}\mathbb{E}_q[k^2],
\end{equation}
where $\mathbb{E}_q$ is the expectation under the hypergeometric law $q_{N,i,M}$ introduced above. The moments of this distribution read
\begin{align}
	\mathbb{E}_q[k]&=i\rho, \\
	\mathbb{E}_q[k^2]&=i\rho(1-\rho)+i^2 \rho^2,
\end{align}
with $\rho=M/N$. We thus obtain
\begin{equation}
	\sum_{k=0}^{\min(i, M)} q_{N,i,M}(k)   \left[ 1 - \sum_{j=1}^k \dfrac{i-j}{N}\right] = 1 - \binom{i}{2}\dfrac{\rho(2-\rho)}{N}.
\end{equation}
Let us now show that the other terms of \cref{eq:Gii_hyper}, coming from the third term and beyond of \cref{eq:expand} are negligible. Note that $i \ll N$, which implies $i(i-1)=\mathcal{O}(1)$. Consider the $l$-th term of \cref{eq:expand}, which is a sum over $j_1,j_2,\dots ,j_l$, where $l \leq k$, and $l\geq 2$ since we already dealt with the case $l=1$. Since the sum runs over $k \leq i$, we have $l \leq i$.
We aim to compute
\begin{equation}
	\sum_{k=0}^{\min(i, M)} q_{N,i,M}(k) \dfrac{1}{l!}\sum_{\substack{j_1,\dots,j_l=1 \\ j_1,\dots,j_l\,\, \textrm{p.d.}}}^k \dfrac{(i-j_1)\dots(i-j_l)}{N^l}.
\end{equation}
This sum comprises $k(k-1)\dots(k-l+1)$ terms, and each of them is bounded as follows:
\begin{equation}
	\dfrac{(i-j_1)\dots(i-j_l)}{N^l} \leq \dfrac{i^l}{N^l}.
\end{equation}
Then, we get
\begin{align}
	\sum_{k=0}^{\min(i, M)} q_{N,i,M}(k) \dfrac{1}{l!}\sum_{\substack{j_1,\dots ,j_l=1 \\ j_1,\dots ,j_l\,\, \textrm{p.d.}}}^k \dfrac{(i-j_1)\dots (i-j_l)}{N^l} &\leq \sum_{k=0}^{\min(i, M)} q_{N,i,M}(k) \dfrac{k(k-1)\dots (k-l+1)}{l!} \dfrac{i^l}{N^l} \nonumber\\
	&\leq \sum_{k=0}^{\min(i, M)} q_{N,i,M}(k) k(k-1)\dfrac{i^{l-2}}{l!} \dfrac{i^l}{N^l}
	\nonumber\\&\leq \dfrac{i^{2l-2}}{l! N^l}\mathbb{E}_q[k(k-1)]\,,
\end{align}
and we use $\mathbb{E}_q[k(k-1)]=\rho^2 i(i-1)$ to obtain
\begin{equation}
	\sum_{k=0}^{\min(i, M)} q_{N,i,M}(k) \dfrac{1}{l!}\sum_{\substack{j_1,\dots ,j_l=1 \\ j_1,\dots ,j_l\,\, \textrm{p.d.}}}^k \dfrac{(i-j_1)\dots (i-j_l)}{N^l} \leq \dfrac{\rho^2 i^{2l}}{N^l l!} \leq \dfrac{\rho}{N} \dfrac{\rho i^{2l}}{N^{l-1}l!}.
\end{equation}
For all $l\geq 2$, this term is negligible with respect to $\rho/N$. Therefore, we have
\begin{equation}
	G_{i,i}= 1-\binom{i}{2}\dfrac{\rho(2-\rho)}{N}+o\left(\frac{\rho}{N} \right). \label{eq:Gii_largeN}
\end{equation}
As expected, this coincides with Kingman's coalescent with the time scaling where one considers the discrete process every $c_N = N/[\rho (2 - \rho)]$ time steps, see \cref{eq:GiiK} and \cref{sec:Kingman}. 
From the general result in \cref{eq:Gii_largeN}, we recover known cases for both Moran and Wright-Fisher models~\cite{gladstien1978}:
\begin{align}
	G_{i,i}\left(\rho = \frac{1}{N}\right)& = 1- \binom{i}{2} \frac{2}{N^2} + o\left(\frac{1}{N^2} \right),
	\\
	G_{i,i}\left(\rho =1\right)& = 1- \binom{i}{2} \frac{1}{N} + o\left(\frac{1}{N} \right).
\end{align}

\paragraph{Large-$N$ limit of $G_{i,i-1}$.}
A similar calculation can be performed for $G_{i,i-1}$, the probability of having one coalescence in one discrete time step, starting from the expression of $G_{i,i-1}$ for finite $N$ given in \cref{eq:Gii-1_gen}. To obtain the large-$N$ limit, it is more straightfoward to use the relation of Gladstien~\cite{gladstien1978} and Felsenstein~\cite{felsenstein1971rate} given in \cref{eq:gladstien}. This equation is valid for $i=1,\dots ,N$. Let us restrict to $i \ll N$ and consider the large-$N$ limit. In that case, we can use \cref{eq:Gii_largeN} to express $G_{i,i}$ and $G_{i-1,i-1}$. This yields
\begin{equation}
	G_{i,i-1} = \left(G_{i-1,i-1}-G_{i,i}\right)\dfrac{i}{2} = \binom{i}{2}\dfrac{\rho(2-\rho)}{N} + o\left(\frac{\rho}{N} \right).
	\label{eq:Gii-1}
\end{equation}
It is interesting to note that
\begin{align}
	G_{i,i-1}& = 1-  G_{i,i} + o\left(\frac{\rho}{N} \right),
	\label{eq:nomorethan2}
\end{align}
which entails that events where more than two lineages coalesce in one time step are negligible with respect to coalescence of two lineages.
Again, we recover the limits of Moran and Wright-Fisher models~\cite{gladstien1978}:
\begin{align}
	G_{i,i-1}\left(\rho = \frac{1}{N}\right)& =  \binom{i}{2} \frac{2}{N^2} + o\left(\frac{1}{N^2} \right),
	\\
	G_{i,i-1}\left(\rho =1\right)& =  \binom{i}{2} \frac{1}{N} + o\left(\frac{1}{N} \right).
\end{align}

\subsection{Coalescence time}

\paragraph{General expression for finite population size $N$.}
Let $T_{i,1}$ be the mean time of coalescence starting with $i$ individuals, i.e.\ the number of time steps elapsed since the time when these individuals last had a common ancestor. It satisfies the recurrence relation
\begin{equation}
	T_{i,1} = 1 + \sum_{j = 1}^i G_{i,j} T_{j,1}= 1 + \sum_{j = 2}^i G_{i,j} T_{j,1}\,,
\end{equation}
and we have $T_{1,1} = 0$. We can rewrite this equation as
\begin{align}
	T_{i,1} &= 1 + \sum_{j = 2}^i G_{i,j} \left[1 + \sum_{k = 2}^i G_{j,k} T_{k,1} \right] = 1 + \sum_{j = 2}^i G_{i,j} \left\{1 + \sum_{k = 2}^i G_{j,k} \left[1 + \sum_{l = 2}^i G_{k, l} T_{l,1} \right] \right\} \nonumber\\
	&= 1 + \sum_{j = 2}^i G_{i,j}  +  \sum_{j = 2}^i \left(G^2\right)_{i,j} +  \sum_{j = 2}^i \left(G^3\right)_{i,j} T_{j,1}\,.
\end{align} 
Note that we have used $\sum_{k = 2}^j G_{j,k}=\sum_{k = 2}^i G_{j,k}$ because $j\leq i$ and $G_{j,l}=0$ if $l>j$. 
We can show by induction that
\begin{align}
	T_{i,1} &= 1 + \sum_{\alpha = 1}^{p} \sum_{j = 2}^i \left(G^{\alpha}\right)_{i,j} + \sum_{j = 2}^i \left({G^{p+1}}\right)_{i,j}T_{j,1}\,,
\end{align} 
for any integer $p\geq 1$. In the limit $p \rightarrow +\infty$, the last term tends to zero, yielding
\begin{align}
	T_{i,1} &= 1 + \sum_{\alpha = 1}^{+\infty} \sum_{j = 2}^i \left({G^{\alpha}}\right)_{i,j}= \sum_{\alpha = 0}^{+\infty} \sum_{j = 2}^i \left({G^{\alpha}}\right)_{i,j} = \sum_{j = 2}^i\left(\mathbf{1}- {G}\right)^{-1}_{i,j}\,.
\end{align} 
where $\mathbf{1}$ refers to the identity matrix. 

Let us now determine the coalescence time, i.e.\ the mean time $T_{\text{MRCA}}$ elapsed since the most recent common ancestor of the whole population. It is defined as 
\begin{align}
	T_{\text{MRCA}} = T_{N,1} = \sum_{j = 2}^N \left(\mathbf{1}- {G}\right)^{-1}_{N,j}\,.
	\label{eq:tmrca_exact}
\end{align} 
This expression allows to obtain the exact formula of $T_{\text{MRCA}}$ for finite $N$, using the general expression of $G_{ij}$ in \cref{eq:Gijgal}. 

\begin{figure}[h!]
	\includegraphics[width = 0.65\columnwidth]{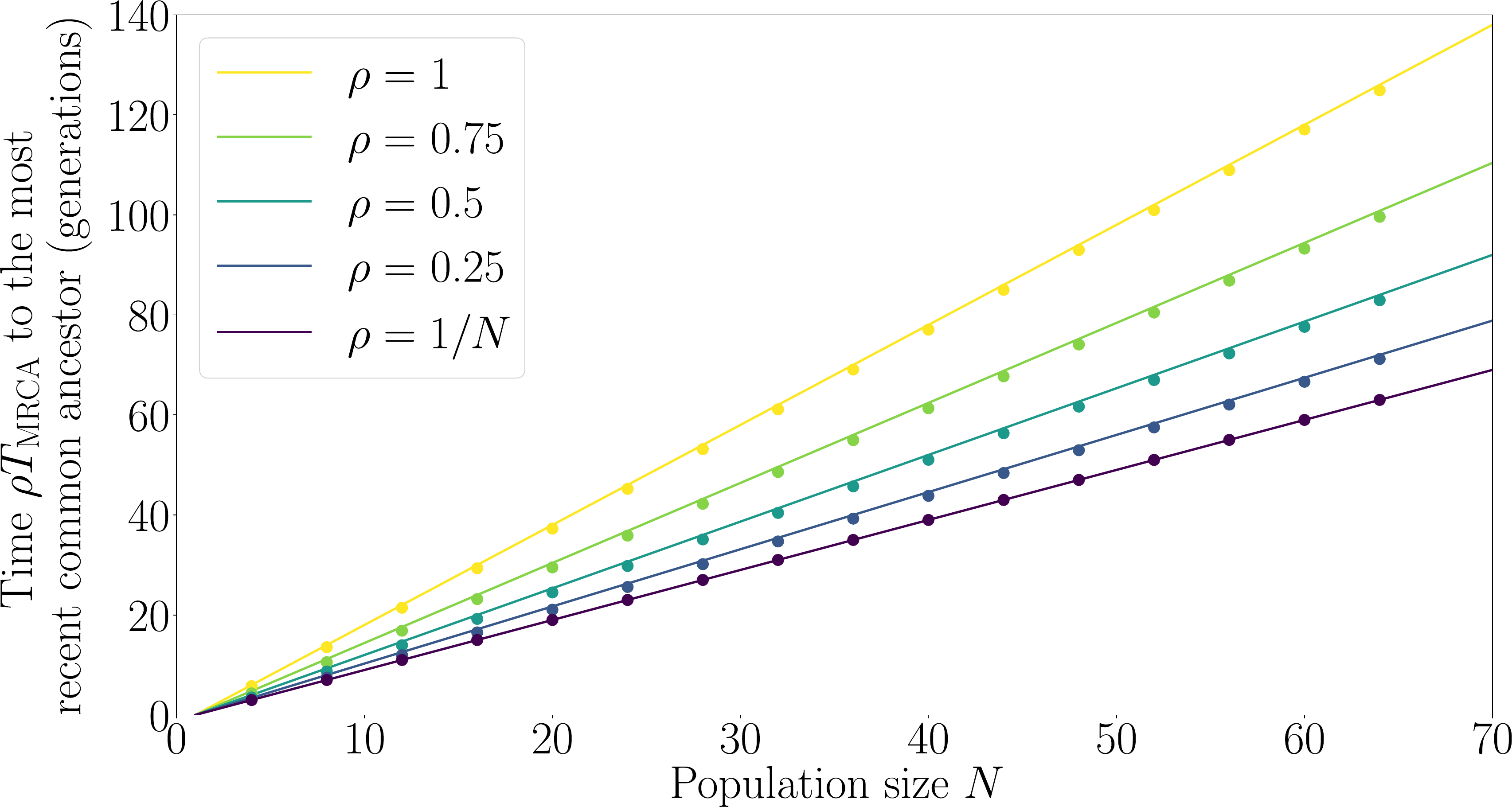}
	\caption{Time $T_{\text{MRCA}}$ to the most recent common ancestor starting with $N$ individuals, expressed in generations (i.e.\ rescaled with the generation time $1/\rho$), as a function of the population size. Markers: numerical calculation of \cref{eq:tmrca_exact}. Lines: analytical result in the large-$N$ limit, given in \cref{eq:T_mrca_large_N}. Exception: for $\rho = 1/N$ (Moran model), the line shows the exact solution $\rho T_{\text{MRCA}} = N-1 $~\cite{ewens1982concept}.}
	\label{fig:t_MRCA}
\end{figure}

\cref{fig:t_MRCA} shows the coalescence time $T_{\text{MRCA}}$ versus population size $N$. Note that while $T_{\text{MRCA}}$ is expressed in numbers of discrete time steps, we show $\rho T_{\text{MRCA}}$ in \cref{fig:t_MRCA}, which is expressed in generations. This allows a meaningful comparison of timescales for models with different updated fractions $\rho$. We observe that the coalescence time is longer in generations for models with larger updated fractions $\rho$.

\paragraph{Large population size limit.}
Let us now focus on the large-$N$ limit.
We start by looking at a sample of size $n$, with $n\ll N$. As discussed in \cref{sec:Kingman}, its genealogy is described by Kingman's coalescent when $N\rightarrow \infty$. 
Since $G_{i,i-1} = 1- G_{i,i}$ in the limit $N\gg 1$ (see \cref{eq:nomorethan2}), the number $\tau_i$ of time steps to move from a number $i\leq n$ of ancestors to a number $i-1$ of ancestors follows the geometric distribution
\begin{equation}
	p(\tau_i = k) = G_{i,i}^{k-1}(1-G_{i,i})\,.
\end{equation}
Thus, in the large-$N$ limit, using \cref{eq:Gii_largeN,eq:Gii-1}, the mean and variance of $\tau_i$ can be expressed as:
\begin{equation}
	\left<\tau_i \right> = \frac{1}{G_{i,i-1}}  = \dfrac{N}{\binom{i}{2} \rho (2-\rho)}\,, 
\end{equation}
Then, as all $\tau_i$ are independent~\cite{kingman1982a, kingman1982b}, we can compute the mean time $T_{n,1}$ to the most recent common ancestor of a sample of size $n$ in the large-$N$ limit as
\begin{align}
	T_{n,1} = \sum_{i=2}^{n} \langle\tau_i\rangle\,.
\end{align}
Thus, in the large-$N$ limit, we obtain (as shown in Refs.~\cite{tavare2004part, ewens2004mathematical})
\begin{equation}
	T_{n,1} = \frac{2N}{\rho(2-\rho)}\left(1-\frac{1}{n}\right)\,.
	\label{eq:Tn_large_N}
\end{equation}
While this derivation only applies to $n \ll N$, it was shown in the general framework of the Cannings model that the average time $T_{\text{MRCA}}$ to reach the most recent common ancestor of the whole population reads in the large-$N$ limit~\cite{ewens2004mathematical, tavare2004part}
\begin{equation}
	T_{\text{MRCA}} = \dfrac{2(N-1)}{\rho(2-\rho)} ,
	\label{eq:T_mrca_large_N}
\end{equation}
which is in fact consistent with our result \cref{eq:Tn_large_N} obtained for $n\ll N$. 
Qualitatively, the reason for this agreement is that the dominant terms, which take more time, are those that regard the coalescence of a small sample in the population.
\cref{fig:t_MRCA} shows that this result is in good agreement with the exact expression we obtained for finite $N$ in \cref{eq:tmrca_exact}, even for rather small values of $N$.
Taking one unit of coalescent time as $c_N=N/[\rho(2-\rho)]$ time steps of the partial update model allows us to recover Kingman's coalescent, which corresponds to $\rho=1$.

\subsection{Coalescent effective population size}
In Ref.~\cite{wakeley2009extensions}, Wakeley and Sargsyan defined the coalescent effective size for neutral models that converge to Kingman's coalescent. They introduced two probabilities that describe the ancestry of the process:
\begin{itemize} 
	\item the probability $\alpha_N$ that a given pair of lineages is descended from a common ancestor at the previous generation,
	\item the probability $\beta_N$ that a single lineage is newly born.
\end{itemize}

In our partial update model, $\beta_N=\rho$, as the probability of being a newly born lineage is simply the probability of being among the $M$ newly produced offspring. We recover the limits $\beta_N=1$ for the Wright-Fisher model and $\beta_N=1/N$ for the Moran model. Let us now compute $\alpha_N$ in our partial update model. We follow Tavaré's argument~\cite{tavare2004part} that applies to Cannings model, and thus in particular to our partial update model in the neutral case. Let us recall that we consider a model for a population of size $N$, where the offspring number in the next step of each individual $i$ is a random variable $\nu_i$, see \cref{sec:Kingman}. From $\nu_1+\cdots+\nu_N = N$, we have $\mathbb{E}[\nu_i]=1$ for all $i$. \cref{eq:Vnu1} states that
\begin{equation}
	\mathbb{V}[\nu_i]=\rho(2-\rho).
\end{equation}
When sampling two random lineages at a given time step, the first one is descended from individual $i$ at the previous step with probability $\nu_i/N$, and the second one is also descended from individual $i$ at the previous step with probability $(\nu_i-1)/(N-1)$. Therefore, the average probability of the two lineages descending from a common ancestor reads
\begin{equation}
	\alpha_N = \mathbb{E} \left[ \sum_i \dfrac{\nu_i(\nu_i - 1)}{N(N-1)}\right]= \dfrac{\mathbb{E}[\nu_i(\nu_i - 1)]}{N-1}=\dfrac{\rho(2-\rho)}{N}+o\left(\frac{\rho}{N}\right).
\end{equation}
This is the inverse of $c_N=N/[\rho(2-\rho)]$, the time rescaling needed to obtain convergence to Kingman's coalescent, see \cref{sec:Kingman}. Note that we recover the limits $\alpha_N=1/N$ for the Wright-Fisher model and $\alpha_N=2/N^2$ for the Moran model~\cite{kingman1982a,wakeley2009extensions}. 

Wakeley and Sargsyan~\cite{wakeley2009extensions} defined the coalescent effective size as $N_{e,c}=\beta_N/\alpha_N$. In our partial update model, this yields
\begin{equation}
	N_{e,c}=\dfrac{\beta_N}{\alpha_N}= \dfrac{N}{2-\rho}.
	\label{eq:effpopsz}
\end{equation}
This result coincides with our results for the other definitions of the effective size we consider here, namely the variance effective population size, and the eigenvalue effective population size (see \cref{sec:effective_pop} of the main text). Let us thus denote it simply by $N_e$.

Note that the probability $\alpha_N$ corresponds to the probability $P_\textrm{ibd}$ of \emph{identity by descent} for a diploid population. Our coalescent approach can be extended to the diploid case by replacing $N$ with $2N$, leading to $P_\textrm{ibd} = \rho(2-\rho)/(2N)$.

\begin{figure}[h!]
	\includegraphics[width = 0.4\columnwidth]{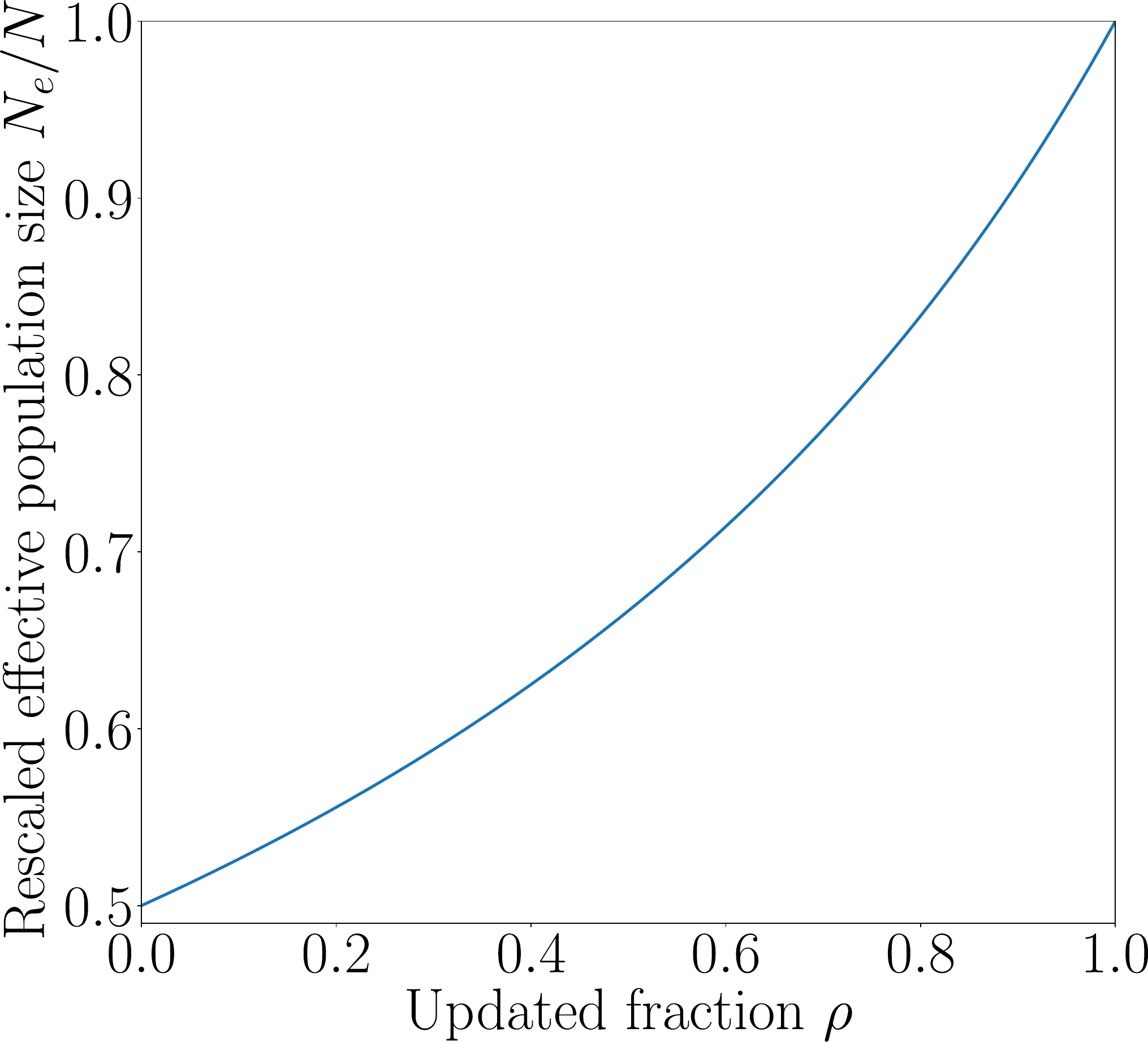}
	\caption{Effective population size $N_e$ (rescaled with $N$) from \cref{eq:effpopsz} as a function of the updated fraction $\rho$.}
	\label{fig:N_e_rho}
\end{figure}

\cref{fig:N_e_rho} shows that the rescaled effective population size $N_e/N$ increases monotonically and non-linearly with the updated fraction $\rho$, from the value $1/2$ corresponding to the Moran model to the value $1$ corresponding to the Wright-Fisher model. 

\newpage
\section{Model with selection on the lifetime: first moments of $\Delta x$}
In this section, we compute the mean and variance of $\Delta x = \Delta X/N$ in a model with selection on the life time of individuals. In this model, we assume that the lifetime of wild-types (resp.\ mutants) is geometrically distributed with parameter $p_w$ (resp.
$p_m = (1 - s)p_w$),  where $s$ denotes the selection
parameter. The transition matrix is given in \cref{eq:P_i_j_selection_life_time} in the main text. Using the same notations as in \cref{sec:diffusion_approx} above, the mean of $\Delta X$ reads
\begin{align}
	\mathbb{E}(\Delta X) & = \sum_{k,m} \binom{i}{k} p_m^k(1-p_m)^{i-k} \binom{N-i}{m-k} p_w^{m-k}(1-p_w)^{N-i-m+k}  \sum_{\Delta X}\binom{m}{k+ \Delta X}x^{k+\Delta X} (1-x)^{m-k-\Delta X} \Delta X \nonumber\\ & = \sum_{k,m} \binom{i}{k} p_m^k(1-p_m)^{i-k} \binom{N-i}{m-k} p_w^{m-k}(1-p_w)^{N-i-m+k} (m x -k) \nonumber\\ & = \sum_{k} \binom{i}{k} p_m^k(1-p_m)^{i-k} \sum_m \binom{N-i}{m-k} p_w^{m-k}(1-p_w)^{N-i-m+k} (m x -k) \nonumber\\ & = \sum_{k} \binom{i}{k} p_m^k(1-p_m)^{i-k} (x p_w (N-i) - k (1- x) ) \nonumber\\ & = x p_w (N-i) - (1-x)p_m i  =N x (1-x) (p_w- p_m)\nonumber\\ &=Np_w sx(1-x)\,.
\end{align}
Meanwhile, the second moment is
\begin{align}
	\mathbb{E}(\Delta X^2) & = \sum_{k,m} \binom{i}{k} p_m^k(1-p_m)^{i-k} \binom{N-i}{m-k} p_w^{m-k}(1-p_w)^{N-i-m+k}  \sum_{\Delta X}\binom{m}{k+ \Delta X}x^{k+\Delta X} (1-x)^{m-k-\Delta X} \Delta X^2 \nonumber\\ & = \sum_{k} \binom{i}{k} p_m^k(1-p_m)^{i-k} \sum_m \binom{N-i}{m-k} p_w^{m-k}(1-p_w)^{N-i-m+k} \left\{m^2 x^2 + m \left[x(1-x) - 2 k x \right] +k^2\right\} \nonumber\\ & = \sum_{k} \binom{i}{k} p_m^k(1-p_m)^{i-k} \left\{ x^2 \left[ (N-i)^2 p_w^2 + (N-i) p_w(1-p_w) \right] + x (1-x) (N-i) p_w (1-2k) \right. \nonumber\\ &  \left. +  k x (1-x) + k^2 (1-x)^2 \right\} \nonumber\\ & = x^2 \left[ (N-i)^2 p_w^2 + (N-i) p_w(1-p_w) \right] + x (1-x) (N-i) p_w  \nonumber\\ &   +  x (1-x) (1-2p_w (N-i) ) i p_m + (1-x)^2 \left[ i^2 p_m^2 + i p_m (1-p_m) \right]\,.
\end{align}

Finally, to first order in $1/N$, the mean and variance of $\Delta x = \Delta X /N$ read
\begin{align}
	\mathbb{E}(\Delta x) & =  p_w s x (1-x)  + o\left( N^{-1}\right)\,, \\ \mathbb{V}(\Delta x) & = \frac{p_w(2- p_w)}{N} x (1-x)   + o\left( N^{-1}\right)\,.
\end{align}

\bibliographystyle{naturemag}

\end{document}